\documentclass[prb,aps,showpacs,superscriptaddress,amsmath,amssymb,floatfix,twocolumn]{revtex4-1}
\usepackage{graphicx}
\usepackage{color}
\usepackage{young}

\begin{document}
\title{
Exact diagonalization of Heisenberg $SU(N)$ chains in the fully symmetric and antisymmetric representations
}
\date{\today} 

\author{Pierre Nataf, Fr\'ed\'eric Mila}
\affiliation{Institut de Physique Th\'eorique, \'Ecole Polytechnique F\'ed\'erale de Lausanne (EPFL), CH-1015 Lausanne, Switzerland}

\begin{abstract}
Motivated by recent experimental progress in the context of ultra-cold multi-color fermionic atoms
in optical lattices, we have developed a method to exactly diagonalize the Heisenberg $SU(N)$ Hamiltonian with
several particles per site living in a fully symmetric or antisymmetric representation of  $SU(N)$.
The method, based on the use of standard Young tableaux, takes advantage of the full $SU(N)$ symmetry, 
allowing one to work directly in each irreducible representations of the global $SU(N)$ group.
Since the $SU(N)$ singlet sector is often much smaller than the full Hilbert space, this enables one to reach 
much larger system sizes than with conventional exact diagonalizations. The method is applied to the study of 
Heisenberg chains in the symmetric representation with two and three particles per site up to $N=10$ and up to
20 sites. For the length scales accessible to this approach, all systems except the Haldane chain ($SU(2)$ with 
two particles per site) appear to be gapless, and the central charge and scaling dimensions extracted
from the results are consistent with a critical behaviour in the $SU(N)$ level $k$ Wess-Zumino-Witten universality
class, where $k$ is the number of particles per site. These results point to the existence of a cross-over between this
universality class and the asymptotic low-energy behavior with a gapped spectrum or a critical behavior in the 
$SU(N)$ level $1$ WZW universality class.
\end{abstract}

     \maketitle

\section{Introduction}
Recent advances in ultracold atoms allow experimentalists to artificially engineer advanced models of strongly correlated systems \cite{dalibard}. In particular, alkaline-earth atoms such a $^{137}Yb$ or $^{87}Sr$ loaded in optical lattices
can be used to realize the Fermi-Hubbard model $SU(N)$ interaction symmetry \cite{gorshkov2010,takahashi2012,pagano2014,scazza,zhang_SUN}. When the number of particle per site $m$ is an integer, and when the on-site repulsion is large enough, the system is expected to be in a Mott insulating phase, which is well described by the Heisenberg $SU(N)$ model. This is a generalization of the familiar $SU(2)$  spin $1/2$ Heisenberg model. 
Depending on the geometry of the lattice (a chain, or the square, triangular, honeycomb.. lattices in 2D), the number of colors $N$, the number of particles per site (and in particular the $SU(N)$ symmetry or - {\it irreducible representation}- of the local wave-function on each site), such a model can lead to a rich variety of quantum phases.
For instance, in $1D$ and for $m=1$, the $SU(N)$ chain, for which a general Bethe ansatz solution exists \cite{sutherland}, is gapless with algebraic decaying correlations, while the same system with $m=2$ can lead to the opening of the gap.
The famous Haldane gap appears for SU(2) with an even number $m$ of particles per site in the totally symmetric representation (corresponding to spin $j=m/2$) \cite{haldanegap,whiteprb1993} .
In $2D$,  for $m=1$, the ground state has been shown to be characterized by some N\'eel-type ordering for $SU(2)$, $SU(3)$\cite{toth2010,bauer_three-sublattice_2012}, $SU(4)$\cite{corbozSU42011} and $SU(5)$\cite{nataf2014} on the square lattice, while the $SU(4)$
model on the honeycomb lattice is an algebraic spin liquid\cite{corbozPRX2012}.
Moreover, on the square lattice, with $m$ particles per site in an antisymmetric representation, the ground state has been predicted by mean-field theory to be a chiral spin liquid provided that $m/N>5$ \cite{hermele2009,hermele_topological_2011}.

From a theoretical point of view, apart from 1D with one particle per site, where the system is both Bethe ansatz solvable \cite{sutherland} and can be studied by a Quantum Monte Carlo algorithm free from the minus sign problem \cite{Frischmuth1999,Messio2012,bonnes}, the study of these systems is in fact often challenging.
Analytical studies can be made with the help of quantum field theory in some large N development\cite{sachdev},  strong coupling limit\cite{lecheminant2015,nonne2011}, or mean-field approach \cite{hermele2009,hermele_topological_2011,szirmaiSU62011}, or through flavor-wave theory \cite{papanicolaou1988,joshi1999}. There is a crucial need to associate those with numerical methods in order to test their validity, or to compensate for them when they are unapplicable or inconclusive.
Among them, Quantum Monte Carlo can be used only in very specific cases (to avoid sign problem): as we already said, in $1D$ for $m=1$, and on any bipartite lattice provided that pairs of interacting sites correspond to conjugate irreducible representations ('irrep')\cite{capponiQMC,assaad2005,cai2013,lang2013,zhou2014}.
In the case where the local wave-function on each site is completely antisymmetric, variational Monte Carlo simulations based on Gutzwiller projected wave-functions have been found to lead to remarkably accurate results \cite{wang_z2_2009,paramekanti_2007,lajko_tetramerization_2013,dufourPRB2015}, but it is not clear how generalize this approach to other irreps, the totally symmetric one for instance.
Density Matrix Renormalization Group (DMRG) methods have also been employed to investigate $SU(N)$ Hamiltonians in $1D$ \cite{rachel2009,nonne2013,quella2012,fuhringer2008,manmana2011}, as well as Infinite Projetced Entangled Pair States ({\it iPEPS}) in $2D$, in a very efficient way \cite{bauer2012,corbozPRX2012,corbozsimplex2012,corbozSU42011}, but the performances of both methods significantly decrease when the dimension of the local Hilbert space increases, as a consequence of the large number of colors $N$ (typically $N\geq6$), or of the large number of particles per site $m$. Finally, the Exact Diagonalization (ED) are limited by the size of the clusters.

Recently, we have developed a method to exactly diagonalize the Hamiltonian for one particle per site independently in each global irrep of $SU(N)$ by using standard Young tableaux\cite{nataf2014}.
Since, for antiferromagnetic interactions, the ground state is in general a singlet, and since the $SU(N)$ singlet sector has a dimension much smaller than that of the full Hilbert space, this enables us to reach essentially the same sizes for a large $N$ as for small $N$:
typically, if we call $N_s$ the number of sites, $N_s\sim 30$ sites.
A natural question was the generalization of the method to larger number of particles per site: $m>1$.

In the present paper, we proceed to this generalization in the cases where the local Hilbert space is a totally symmetric or antisymmetric irrep.
In the first part, we explain our method which is also based on the use of standard Young tableaux.
We build an orthonormal basis of states belonging to each irrep of $SU(N)$, and show how to write the $SU(N)$ two-sites interaction in such a basis.
Then, we apply this method to the study of the Heisenberg $SU(N)$ symmetric chain with $m=2$ and $m=3$ particles per site, in order to investigate the problem of the Haldane gap in the context of $SU(N)$ chain\cite{rachel2009}, a problem still open for most values of $N$ ($N\geq4$) in spite of the efficiency of DMRG algorithm to treat 1D short range interactions Hamiltonian. The results turn out to be quite surprising: for all systems except the Haldane chain ($SU(2)$ with two particles per site), the excitation gap seems to tend to zero with the system size, consistent with a gapless behavior. In addition, the central charge and the scaling dimension that could be extracted from the finite-size energies are consistent  with the $SU(N)$ level $k$ WZW universality class, where $k$ is the number of particles per site (then $k=m$). These results contradict the DMRG results of Ref.~\onlinecite{rachel2009} for $N=3$ and the field theory
expectation that, if the system is critical, the universality class should be $SU(N)$ level $1$ WZW. We propose an explanation in
terms of a cross-over between $SU(N)$ level $k$ WZW at intermediate energies, and a gapped behavior or $SU(N)$ level $1$ WZW at low energy.

\begin{figure}
\begin{center}
\includegraphics[width=230pt]{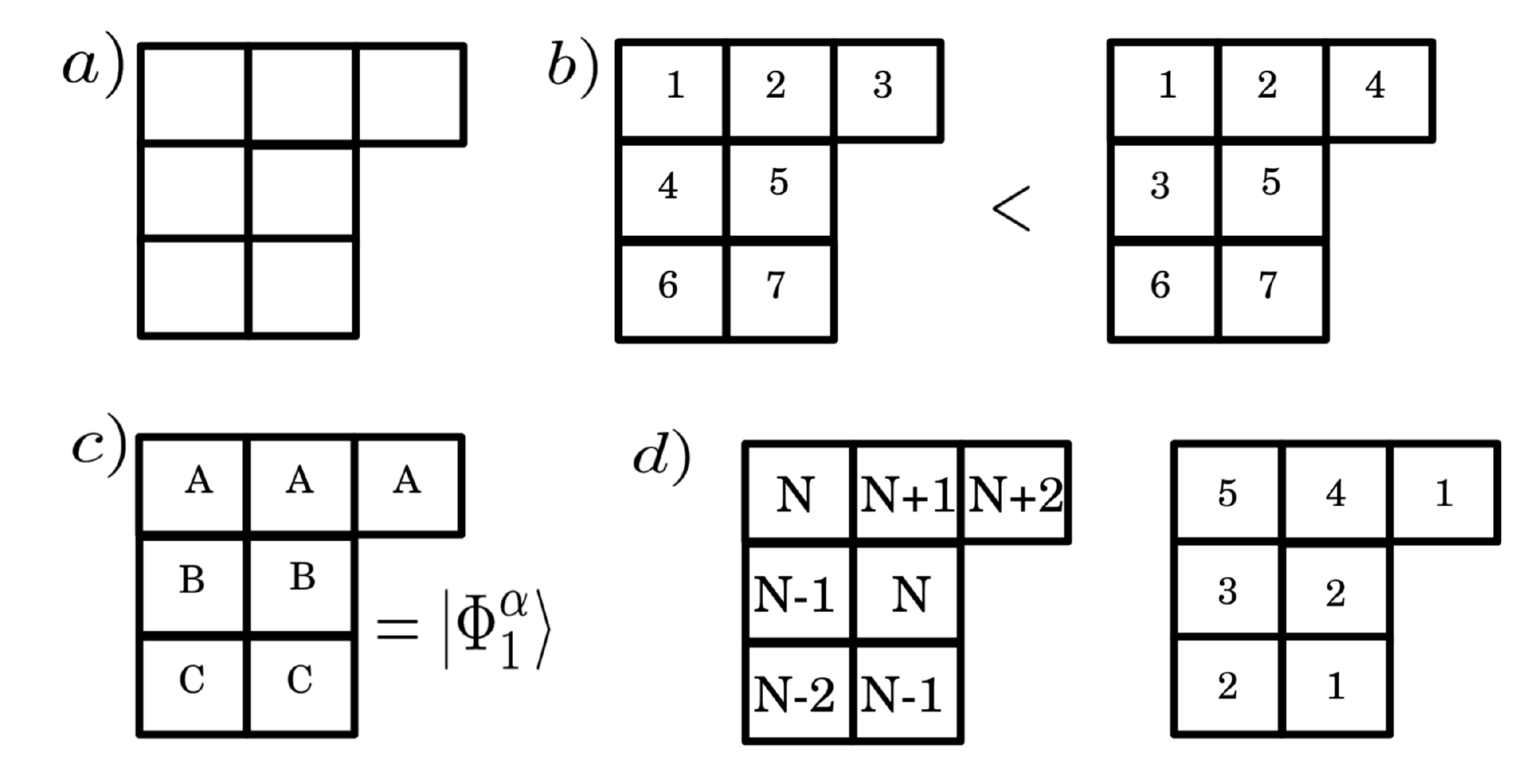}
\vskip -0.2cm
\caption{\label{schema_short}a) Young tableau of shape $[3,2,2]$; b)examples of standard tableaux ranked according to the {\it last letter sequence}; c) $|\Phi_1^{[3,2,2]}\rangle=|AAABBCC\rangle$; d) left: integers $d_{i,N}$ that enter into the calculation of the dimension $d^{\alpha}_N$; right: hook lengths $l_i$; $d_{i,N}$ is the product of the numbers of the left box divided by the numbers of the right box.
}
\vspace{-0.5cm}
\end{center}
\end{figure}

\section{The method}
In the most general case, a  $SU(N)$ Heisenberg-like interaction between two sites $i$ and $j$ can be written as:
\begin{align}
\label{interaction_ij}
H_{(i,j)}= \sum_{\mu, \nu}\hat{S}^{i}_{\mu\nu}\hat{S}^{j}_{\nu\mu} ,
\end{align}
where the $SU(N)$ generators  satisfy on each site $i$ the
following commutation relation:
\begin{equation*}
	\big{[}\hat{S}_{\alpha\beta}^{i},\hat{S}_{\mu\nu}^{i} \big{]} = \delta_{\mu\beta}\hat{S}_{\alpha\nu}^{i}-\delta_{\alpha\nu}\hat{S}_{\mu\beta}^{i}.
\end{equation*}

\subsection{Brief review for one particle per site}
\label{m=1}
When there is one particle per site, the local states belong to the fundamental representation of $SU(N)$.
The local Hilbert space is $N-$dimensional and spanned by $N$ states, one for each color, that we can call A,B,C, $etc$.
The  interaction in Eq.(\ref{interaction_ij}) then takes the form of a permutation operator $P_{i,j}$:
\begin{align}
\label{interaction_ij_oneparticle}
H_{(i,j)}=P_{i,j} ,
\end{align}
which switches the state between site $i$ and $j$: $P_{i,j} |\gamma\rangle_i \otimes  |\beta\rangle_j=|\beta\rangle_i \otimes  |\gamma\rangle_j$,
for any $\gamma, \beta=A,B,C$...
In that case, an efficient method has been devised to work directly in the irreps of the global $SU(N)$ symmetry in Ref.~\onlinecite{nataf2014}.  Here, we just summarize the most important results, and we introduce
the basic definitions needed to understand the rest of the section.

Each irrep of $SU(N)$ is labeled by a Young tableau  $\alpha=[\alpha_1,\alpha_2,...,\alpha_k]$  ($1\leq k \leq N$) where the lengths of the rows $\alpha_j$ satisfy $\alpha_1\geq\alpha_2\geq...\geq \alpha_k \geq 1$ (see Fig.\ref{schema_short} a). The construction relies on the concept of standard Young tableaux
(SYT) associated to a given shape $\alpha$, i.e. tableaux filled with numbers from 1 to $N_s$ (equal to the number of boxes) in ascending order from left to right and from top to bottom.
Their number is denoted by $f^\alpha$, and they can be  ranked from 1 to $f^{\alpha}$ according to the {\it last letter sequence}: two SYTs $S_r$ and $S_s$ are such that $S_r<S_s$ if the number $N_s$ appears in $S_r$ in a row below the one it appears in $S_s$. If those rows are the same, one looks at the rows of $N_s-1$, etc  (see Fig.\ref{schema_short} b).

Then, for a given Young tableau, it has been shown in Ref. ~\onlinecite{nataf2014} that one can construct an orthonormal basis with the help of
linear superposition of permutations $\{o^{\alpha}_{rs} \}_{r,s=1..f^{\alpha}} $ called {\it orthogonal units} which satisfy the property:
\begin{align}
o^{\alpha}_{rs}o^{\beta}_{uv}=\delta^{\alpha \beta } \delta_{su} o^{\alpha}_{rv} \hspace{0.3cm} \forall\,\, r,s=1...f^{\alpha}\,,\forall\, u,v =1...f^{\beta} \label{relation_orthonormal},
\end{align} 
which allow one to write the projector on the irrep $\alpha$ as:
 \begin{equation}
T^{\alpha}=\sum_{r=1...f^{\alpha}}o^{\alpha}_{rr},
\end{equation}
and which,  more generally, allow one to uniquely express any linear superposition of permutations $\eta$: 
\begin{equation}
\eta=\sum_{\beta,t,q} \mu^{\beta}_{tq} (\eta) o^{\beta}_{tq},
\end{equation}
where $\mu^{\beta}_{tq} (\eta)$ are the coefficients of the decomposition.
Indeed, attaching a site to each integer of the SYTs and interpreting the permutations as operators acting in the Hilbert space, the family of states
\begin{align}
\label{basis}
 \Big{\{} |\Psi^{\alpha}_r\rangle =|| o^{\alpha}_{11} |\Phi^{\alpha}_1\rangle ||^{-1} o^{\alpha}_{r1} |\Phi^{\alpha}_1\rangle \Big{\}}_{r=1...f^{\alpha}} 
\end{align}
where $|\Phi^{\alpha}_1\rangle$ is a product state with $A$ on the first line, $B$ on the second line, etc.,(see Fig. \ref{schema_short} c) )
can be proven to be an orthonormal basis of one of the sectors of the irrep $\alpha$ (if the quadratic Casimir of $\alpha$ is not equal to zero, $\alpha$
can be decomposed into equivalent sectors). Most importantly, the matrix $\{\mu^{\alpha}_{tq} (P_{k,k+1})\}_{t,q}$ describing $P_{k,k+1}$, the permutation between neighboring sites $k$ and $k+1,$ takes a very simple form in this basis: if $k+1$ and $k$ are in the same row (resp. column) in $S_t$,
then $\mu^{\alpha}_{tt} (P_{k,k+1})=+1$ (resp. $-1$), and all other matrix elements involving $t$ vanish.
If $k+1$ and $k$ are not in the same column or the same line, and if $S_u$ is the tableau obtained from $S_t$ 
by interchanging $k$ and $k+1$, then the only non-vanishing matrix elements involving $t$ or $u$ are given by:
\begin{align}
\begin{pmatrix}
\mu^{\alpha}_{tt} (P_{k,k+1})&\mu^{\alpha}_{tu} (P_{k,k+1}) \\
 \mu^{\alpha}_{ut} (P_{k,k+1})&\mu^{\alpha}_{uu} (P_{k,k+1}) \end{pmatrix}=\begin{pmatrix} 
-\rho&\sqrt{1-\rho^2} \\ 
 \sqrt{1-\rho^2} &\rho\end{pmatrix} 
 \nonumber
 \end{align}
where
$\rho$ is the inverse of {\it the axial distance} from $k$ to $k+1$ in $S_t$ defined by counting $+1$ (resp. $-1$) for each step made downwards or to the left  (resp. upwards or to the right) to reach $k+1$ from $k$.
Since any permutation can be written as product of permutations between neighboring sites, this allows one to write down very simply the matrix of
the Hamiltonian of Eq.~(\ref{interaction_ij_oneparticle}), which can then be diagonalized using Lanczos algorithm. 

\subsection{The general Hamiltonian as a sum of permutations}

Before entering an explicit construction for symmetric and antisymmetric irreps, let us summarize the main idea of the extension to the general case.
If we have $m$ particles per site, the Hilbert space that corresponds to a specific irrep at each given site is the subspace obtained by applying the appropriate projector at each site to a much larger Hilbert space, where each particle would be in any of the $N$ states. This latter is the same as the Hilbert space for $mN_s$ sites and the fundamental representation at each site. In that Hilbert space, the Hamiltonian of Eq. (~\ref{interaction_ij}) just corresponds to coupling all particles at site $i$ to all particles at site $j$.  
So, by numbering each of the $m$ particles located in each site, it is possible to express this Hamiltonian as a sum of $m^2$  permutations.
Assigning number $m(j-1)+l$ to the $l^{th}$ particle of site $j$, for $l=1...m$, it takes the form (see also Appendix \ref{fermionic}):
 \begin{align} 
\label{interaction}
H_{(i,j)}=\sum^{l'=1...m}_{l=1...m} P_{m(i-1)+l, m(j-1)+l'}.
\end{align}
If we solve this Hamiltonian in the full Hilbert space, the spectrum will include all the spectra obtained with all possible combinations of local irreps.

To work in a specific irrep at each site, one needs to construct an appropriate basis of the projected Hilbert space. Since the Hamiltonian with
one particle per site takes a very simple form in the basis of SYTs, the natural idea is to try and express the projectors in this basis to get a basis
as linear combinations of SYTs. In the following, we show that for the fully symmetric or fully antisymmetric representations, on which we
want to focus in this paper, this can be achieved quite easily. 

\subsection{Symmetric and antisymmetric local irreps}

First, let us give practical examples of the kind of states that {\it live} in such local Hilbert spaces.
If  $m=2$, the fully antisymmetric irrep  labeled by the Young tableau $\lambda=[1,1]$, is represented as:
\begin{align}
\begin{Young}
\cr
\cr
\end{Young}
\end{align}
If $N=4$, and if we call $A,B,C$ and $D$ the four colors, the $6$ orthogonal basis states of the irrep $[1,1]$ can be chosen
as :
\begin{align}
\label{antisymmetric_states}
\{ \frac{1}{\sqrt{2}}\{|AB\rangle-|BA\rangle\}&,\frac{1}{\sqrt{2}}\{|AC\rangle-|CA\rangle\} \nonumber\\
\frac{1}{\sqrt{2}}\{|AD\rangle-|DA\rangle\}&,\frac{1}{\sqrt{2}}\{|BC\rangle-|CB\rangle\} \nonumber\\
\frac{1}{\sqrt{2}}\{|BD\rangle-|DB\rangle\}&,\frac{1}{\sqrt{2}}\{|CD\rangle-|DC\rangle\} \}.
\end{align}
The fully symmetric irrep $\lambda=[2]$ represented as 
\begin{align}
\begin{Young}
& \cr
\end{Young}
\end{align}
is spanned, for $N=4$ by the $10$ states:
\begin{align}
\label{symmetric_states}
\{ |AA\rangle, |BB\rangle&,|CC\rangle,|DD\rangle\nonumber\\
\frac{1}{\sqrt{2}}\{|AB\rangle+|BA\rangle\}&,\frac{1}{\sqrt{2}}\{|AC\rangle+|CA\rangle\} \nonumber\\
\frac{1}{\sqrt{2}}\{|AD\rangle+|DA\rangle\}&,\frac{1}{\sqrt{2}}\{|BC\rangle+|CB\rangle\} \nonumber\\
\frac{1}{\sqrt{2}}\{|BD\rangle+|DB\rangle\}&,\frac{1}{\sqrt{2}}\{|CD\rangle+|DC\rangle\} \}
\end{align}
The dimension $d^{\alpha}_N$ of a (local) $SU(N)$ irrep of shape $\alpha$ can be calculated 
very simply from the shape $\alpha$ as $d^{\alpha}_N=\prod_{i=1}^{n} (d_{i,N}/ l_i )$, with $d_{i,N}=N+\gamma_i$, where  $\gamma_i$ is the algebraic distance from the $i^{th}$ box to the main diagonal, counted positively (resp. negatively) for a box above (below) the diagonal  (see Fig.\ref{schema_short} d). The hook lengths $l_i$ of a box are defined as the number of boxes on the same row at the right plus the number of boxes in the same column below plus the box itself (see Fig.\ref{schema_short} d).
Applying those rules to the fully antisymmetric (resp. symmetric) shapes with $m$ boxes in one column (resp. row) allows one
to obtain the following formulas  (with $\epsilon=+1$ for symmetric and $\epsilon=-1$ for antisymmetric):
\begin{equation}
d^{\epsilon,m}_{N}= \frac{N(N+\epsilon)...(N+\epsilon(m-1))}{m!}.
\label{dim_sym}
\end{equation}
This number can be very high even for small numbers of particles and could prohibit the diagonalisation of the Hamiltonian 
even on small clusters since the full Hilbert space has dimension $(d^{\epsilon,m}_{N})^{N_s}$, where $N_s$ is the number of sites.
For instance, for $SU(10)$ and $m=2$ in the symmetric irrep, $d^{+,2}_{10}=55$.
Yet, we have been able, thanks to the method we present below, to find the exact ground state energy for the antiferromagnetic Heisenberg model on the $20$ sites chain for this system,
while $(d^{+,2}_{10})^{20}\simeq 6.42 \times 10^{34}$.

\subsubsection{Selection of relevant symmetric and antisymmetric SYTs}
\label{selection_relevant}
The first important observation (which will be proven in Appendix \ref{proof_1}) is that a SYT will give 0 after projection onto a symmetric or antisymmetric irrep unless the particles at any given site satisfy a simple symmetry condition:
for the antisymmetric case, the numbers corresponding to a given site should be in different rows, while for the symmetric
case, the numbers corresponding to a given site should be in different columns. Such SYTs are said to be relevant. 

For instance, for $m=2$ and $N_s=4$, the following SYT is relevant for the antisymmetric case:
\begin{align}
{\ensuremath \raisebox{-16pt}{$ \begin{Young}
1&3\cr
2&5\cr
4&7\cr
6&8 \cr
\end{Young} $}}
\end{align}
while the following one is not:
\begin{align}
{\ensuremath \raisebox{-16pt}{$ \begin{Young}
1&4\cr
2&5\cr
3&6\cr
7&8 \cr
\end{Young} $}}
\end{align}
since the row of '8' which is the number of the second particle of the fourth site is in the same row as '7', which is the number of the first particle of the fourth site.

\subsubsection{Equivalence classes and representatives}
\label{equivalent_class}
If two relevant SYTs only differ by permutations among the particles of given sites, we say that they 
belong to the same equivalence class. For instance, for 6 sites with $m=2$ particles per site, in the case where $N\geq 4$, the two SYTs
\begin{align}
{\ensuremath \raisebox{-16pt}{$ \begin{Young}
1&2&5\cr
3&4&7\cr
6&9&10\cr
8&11& 12\cr
\end{Young} $}} \,\,\,\,{\ensuremath \raisebox{-16pt}{$ \begin{Young}
1&2&6\cr
3&4&7\cr
5&9&10\cr
8&11& 12\cr
\end{Young} $}}
\end{align}
belong to the same class for the symmetric case because each pair of numbers $2(k-1)+1$ and $2k$ belong to the same pair of locations in the two SYTs. 

The second important observation is that all the SYTs belonging to the same equivalence class lead, after projection, to the same (non vanishing) linear combination of permutations. This can be shown using some properties of the orthogonal units and a counting argument based on the Itzykson-Nauenberg rules (cf Ref \onlinecite{itzykson} and Appendix \ref{proof_2}).

Then, we need to keep only one state per class, that we will call a {\it representative}. To select one representative in each class of SYTs, 
one can for instance use the classification of the
last letter sequence and pick the smallest state. An algorithm that allows to construct this family of SYTs is presented in Appendix \ref{appendix_algo_SYTs}.
To proceed further, we need to specify the form of the projector, hence to work separately for the antisymmetric and the symmetric cases.

\subsubsection{The equivalence classes in the antisymmetric case}
\label{eq_classes_as}
Generally, for a given shape $\alpha$ with $N_sm$ boxes, there are $\tilde{f}^{\alpha} \leq f^{\alpha} $ equivalence classes and representatives. 
We denote the representatives as $\tilde{S}_r$  for $1\leq r \leq \tilde{f}^{\alpha}$, classified according to the last letter sequence.
Due to the selection rules established in the previous paragraph, the $\tilde{f}^{\alpha}$ representatives are  SYTs of shape $\alpha$ with additional {\it internal} constraints:
if we call $y(q)$ the row (between $1$ and $N$) where the number $1\leq q\leq mN_s$ is located in the considered SYT, one must have
$y(m(k-1)+1)<y(m(k-1)+2)<...<y(mk)$ (where $1\leq k\leq N_s$ is the index of the site).
Unfortunately, we are not aware of the equivalent of the {\it hook length formula} to calculate directly the number $\tilde{f}^{\alpha}$ from its shape $\alpha$ \footnote{In paragraph \ref{symmetric_classes}, we give a mathematical definition of those numbers in terms of {\it Kostka} numbers }.
However, and indeed certainly more importantly in view of doing computational calculations, we have an efficient way to generate all the $\tilde{f}^{\alpha}$ SYTs with proper internal constraints for a shape $\alpha$
given as an imput (see Appendix \ref{appendix_algo_SYTs}).
Moreover, one can perform the following decomposition of the full Hilbert space: $\bigotimes_{i=1}^{N_s}(\mathcal{H}\mathcal{S}_i)= \oplus_{\alpha} V^{\alpha}$.
When $V^{\alpha}$ stands for the $SU(N)$ singlets collective irrep ($\alpha$ being a rectangle of dimension $N\times \frac{N_s m}{N}$), the dimension of   $V^{\alpha}$ (number of independent $SU(N)$ singlets) is directly $\tilde{f}^{\alpha}$.
And when $\alpha$ stands for an $SU(N)$ irrep with strictly positive quadratic Casimir, exactly as in the $m=1$ case \cite{nataf2014}, $V^{\alpha}$ can itself be decomposed into $d^{\alpha}_N$ equivalent subsector on which the Hamiltonian is invariant. Each of them has the dimension  $\tilde{f}^{\alpha}$. 
Thus, the decompostion of the full Hilbert space leads to the following equality for the dimensions:
\begin{align}
\label{dimension_fullHS_antisymmetric}
\text{dim}(\bigotimes_{i=1}^{N_s}(\mathcal{H}\mathcal{S}_i))=\text{dim}(\mathcal{H}\mathcal{S}_i)^{N_s}=(d^{-,m}_{N})^{N_s}=\sum_{\alpha} \tilde{f}^{\alpha} d^{\alpha}_N,
\end{align}
where $\alpha$ stands for all the shapes of $N_s m$ boxes and no more than $N$ rows.
Importantly, such an equality can be straightforwardly (and independently) obtained from the Itzykson-Nauenberg rules that we review in Appendix \ref{itzykson_rules}.

\subsubsection{The basis states in the antisymmetric case}
\label{states_as}
Now, to build basis states for a given shape $\alpha$, we just need to apply a projection operator $\mathcal{P}roj$ to the orthogonal units $o^{\alpha}_{r1}$ of the
representatives:
\begin{equation}
\mathcal{P}roj=\prod\limits_{k=1}^{N_s} \mathcal{P}roj(k),
\end{equation}
where $\mathcal{P}roj(k)$ imposes the local antisymmetry at site k:
\begin{equation}
\label{projk}
 \mathcal{P}roj(k)=\frac{1}{m!}\sum_{\sigma\in \mathcal{S}_m(k)} \epsilon(\sigma) \sigma.
\end{equation}
In the last equation, the sum runs over a group that can be named $\mathcal{S}_m(k)$, which gathers all the permutations $\sigma$ that interexchange between each other the $m$ particles of the site $k$, whose numbers are $m(k-1)+1,....,mk$.
The function $\epsilon(\sigma)$ is the signature of the permutation $\sigma$. It is equal to $+1$ (resp. $-1$) for an even (resp. odd)
permutation $\sigma$.
Thus, for instance,  for the site $k=1$, if $m=2$, we have:
\begin{equation}
 \mathcal{P}roj(1)=\frac{1}{2}(\mathcal{I}_d-(1,2)),
\end{equation}
while if $m=3$:
\begin{equation}
 \mathcal{P}roj(1)=\frac{1}{6}\big\{\mathcal{I}_d-(1,2)-(1,3)-(2,3)+(1,2,3)+(1,3,2)\big\},
\end{equation}
where $\mathcal{I}_d$ is the identity, $(1,2)=P_{1,2}$ is the permutation $1\leftrightarrow2$, and so on.\\

Then, the desired set of states can be defined as:
\begin{align}
\label{basis_antisym}
 \Big{\{} |\Psi^{\alpha}_r\rangle =\mathcal{N}^{-1}\mathcal{P}roj \times \frac{ o^{\alpha}_{r1} |\Phi^{\alpha}_1\rangle}{|| o^{\alpha}_{11} |\Phi^{\alpha}_1\rangle ||} \Big{\}}_{r=1...\tilde{f}^{\alpha}} ,
\end{align}
where $\mathcal{N}$ is some normalization constant, and where the index $r$ runs over the representatives SYTs $\tilde{S}_r$  for $1\leq r \leq \tilde{f}^{\alpha}$. Note that  $\frac{ o^{\alpha}_{r1} |\Phi^{\alpha}_1\rangle}{|| o^{\alpha}_{11} |\Phi^{\alpha}_1\rangle ||}$ is a normalized $N_sm$-particles state of $SU(N)$ symmetry $\alpha$ , 
that has no defined property of local symmetry, i.e it appears in the Hilbert space of a system of $N_sm$ sites with one particle per site.
First of all, due to the rules reviewed in section
\ref{m=1}, $\mathcal{P}roj \times  o^{\alpha}_{r1}$ is a linear superposition of $o^{\alpha}_{p(r)1}$, where the indices
$p(r)$ designate SYTs belonging to the same equivalence class as $\tilde{S}_r$.

It implies that for two representatives SYTs $\tilde{S}_r,  \tilde{S}_{r'}$ ($1\leq r<r'\leq \tilde{f}_{\alpha}$),
$\langle\Psi^{\alpha}_{r'}|\Psi^{\alpha}_r\rangle=\delta_{r,r'}$since the two classes of   $\tilde{S}_r$ and  $\tilde{S}_{r'}$ are disjoint.
It also implies that each state $|\Psi^{\alpha}_r\rangle$ belongs to the sector of global $SU(N)$ symmetry $\alpha$.
Finally, since $\forall j=1...N_s$:
 \begin{align} 
P_{m(j-1)+l, m(j-1)+g} \mathcal{P}roj(j)= -\mathcal{P}roj(j)\,\,\,\,\forall 1 \leq l<g\leq m , 
\end{align}
the local antisymmetry is satisfied.

From a conceptual point of view, we could stop here, but we want to give additional details to make the actual implementation easier.
If one identifies each state $ \frac{ o^{\alpha}_{p1} |\Phi^{\alpha}_1\rangle}{|| o^{\alpha}_{11} |\Phi^{\alpha}_1\rangle ||}$ with the corresponding tableau $S_p$, one can express the normalizing projection operator $\mathcal{N}^{-1}\mathcal{P}roj$ as an operator on the SYTs , for $1\leq r\leq \tilde{f}^{\alpha}$:
\begin{align}
\label{tableau_tilde}
\mathcal{N}^{-1}\mathcal{P}roj \frac{ o^{\alpha}_{r1} |\Phi^{\alpha}_1\rangle}{|| o^{\alpha}_{11} |\Phi^{\alpha}_1\rangle ||}\equiv \mathcal{N}^{-1}\mathcal{P}roj \tilde{S}_r =\big{(}\prod \limits_{j=1}^{N_s} \varsigma_j \big{)} \tilde{S}_r ,
\end{align}
where $\varsigma_j$ is a superposition of $m!$ operators that interexchange between each other the numbers $m(j-1)+1,\,\,m(j-1)+2\, ...,\,mj$ in the SYT tableau $\tilde{S}_r$.
 According to the rules controling the effect of successive transposition reviewed in \ref{m=1} and to the definition of the normalized projector operator (cf Eq. (\ref{projk})), it is easy to see that for $m=2$, $\varsigma_j$ must be:
\begin{align}
\label{proj_2part}
\varsigma_j=\sqrt{\frac{1+\rho(j)}{2}}\mathcal{I}_d-\sqrt{\frac{1-\rho(j)}{2}}\mathcal{T}(j),
\end{align}
where $\rho(j)$ is the inverse of the axial distance between the numbers $2j-1$ and $2j$ (which is necessarly non negative due to the internal constraints on the tableau $\tilde{S}_r$),
$\mathcal{I}_d$ is defined as the identity operator on the SYT, while $\mathcal{T}(j)$ switches  $2j-1$ and $2j$ in the tableau on which it is applied.
For example, one has:
\begin{align}
\mathcal{T}(2) \,{\ensuremath \raisebox{-22pt}{$ \begin{Young}
1&3\cr
2&5\cr
4&7\cr
6&8 \cr
\end{Young} $}}  = {\ensuremath \raisebox{-22pt}{$ \begin{Young}
1&4\cr
2&5\cr
3&7\cr
6&8 \cr
\end{Young} $}}\,\,.
\end{align}
Note that in case where $2j-1$ and $2j$ are in the same column (necessarly one above the other),
then $\rho(j)=1$ and $\varsigma_j=\mathcal{I}_d$: the constraint $P_{2j-1,2j}$ directly gives $-1$ on such a SYT.

\begin{figure}
\begin{center}
\includegraphics[width=250pt]{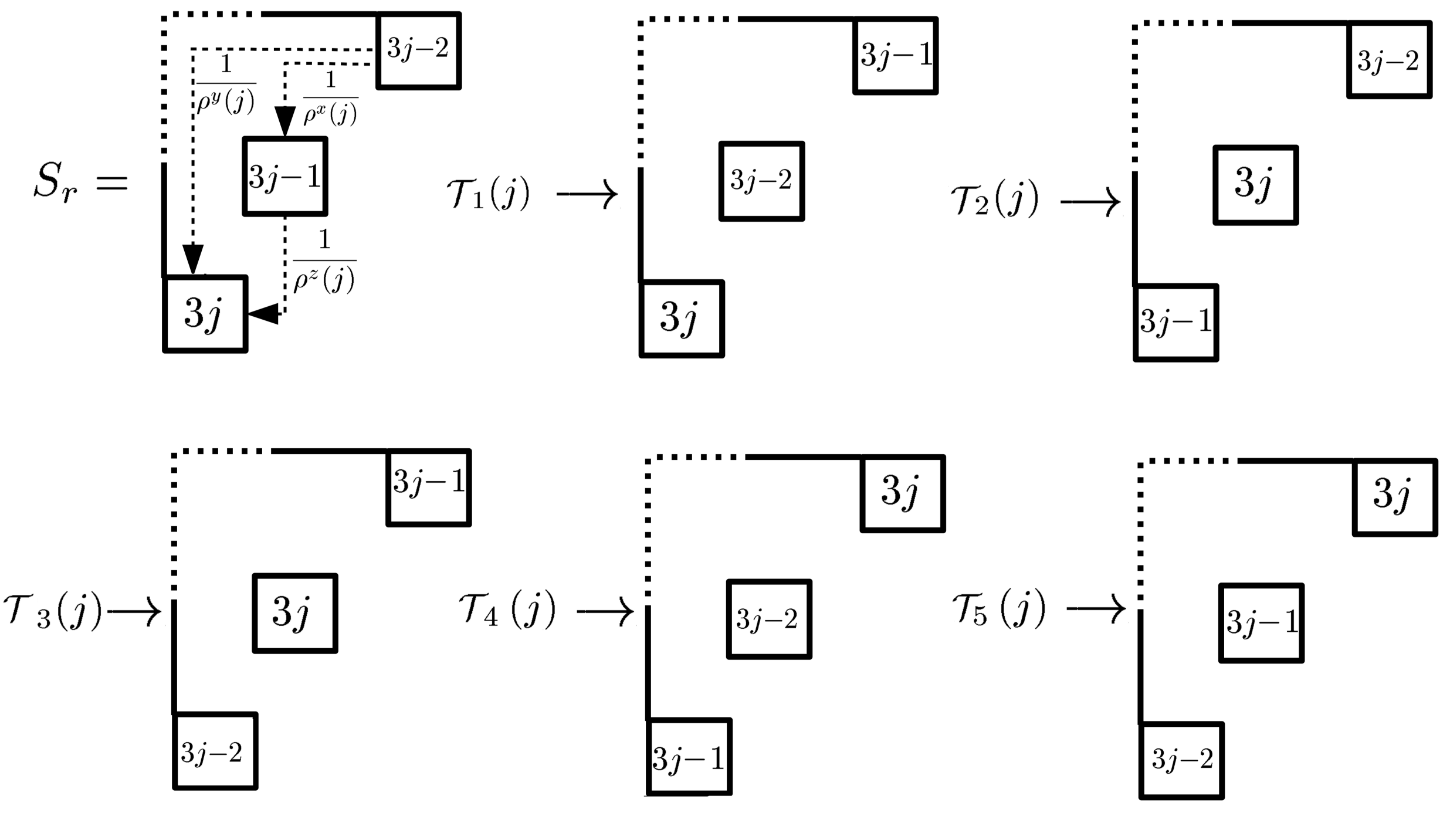}

\caption{\label{3mschema}Action of the operators $\mathcal{T}_q(j)$ on a SYT.
}
\vspace{-0.5cm}
\end{center}
\end{figure}

Thus, for four sites with $m=2$ particles per site, in the case where $N\geq 4$, one state of the irrep $\alpha=[2,2,2,2]$ is for instance:
\begin{align}
\mathcal{N}^{-1}\mathcal{P}roj \,\, {\ensuremath \raisebox{-20pt}{$ \begin{Young}
1&3\cr
2&5\cr
4&7\cr
6&8 \cr
\end{Young} $}} 
\end{align}
which is equal to the superposition:
\begin{align}
 \frac{2}{3}\, {\ensuremath \raisebox{-16pt}{$ \begin{Young}
1&3\cr
2&5\cr
4&7\cr
6&8 \cr
\end{Young} $}} - \frac{\sqrt{2}}{3}\,{\ensuremath \raisebox{-16pt}{$ \begin{Young}
1&4\cr
2&5\cr
3&7\cr
6&8 \cr
\end{Young} $}} - \frac{\sqrt{2}}{3} \, {\ensuremath \raisebox{-16pt}{$ \begin{Young}
1&3\cr
2&6\cr
4&7\cr
5&8 \cr
\end{Young} $}} +\frac{1}{3} \, {\ensuremath \raisebox{-16pt}{$ \begin{Young}
1&4\cr
2&6\cr
3&7\cr
5&8 \cr
\end{Young} $}}\,\,\,,
\end{align}
which is obviously normalized.
For $m=3$, the operators $\varsigma_j$ are a bit more complicated:
\begin{align}
\varsigma_j=\sum \limits_{q=0}^{5} \eta_q(j) \mathcal{T}_q(j).
\end{align}
The vector of coefficients $(\eta(j))_{q=0...5}$ reads:
\begin{align}
\label{coeffantisymm}
\eta(j) = \frac{1}{\sqrt{6}} \begin{pmatrix} \sqrt{1+\rho^x(j)}\sqrt{1+\rho^y(j)}\sqrt{1+\rho^z(j)}\\ 
-\sqrt{1-\rho^x(j)}\sqrt{1+\rho^y(j)}\sqrt{1+\rho^z(j)}\\
-\sqrt{1+\rho^x(j)}\sqrt{1+\rho^y(j)}\sqrt{1-\rho^z(j)}\\
\sqrt{1+\rho^x(j)}\sqrt{1-\rho^y(j)}\sqrt{1-\rho^z(j)}\\
\sqrt{1-\rho^x(j)}\sqrt{1-\rho^y(j)}\sqrt{1+\rho^z(j)}\\
-\sqrt{1-\rho^x(j)}\sqrt{1-\rho^y(j)}\sqrt{1-\rho^z(j)}\\
\end{pmatrix},
\end{align}
where $\rho^x(j)$ is the inverse of the axial distance from the number $3j-2$ to $3j-1$,  $\rho^y(j)$ is the inverse of the axial distance from the number $3j-2$ to $3j$,
and $\rho^z(j)$ is the inverse of the axial distance from the number $3j-1$ to $3j$.
Note that by definition $\frac{1}{\rho^x(j)}+\frac{1}{\rho^z(j)}=\frac{1}{\rho^y(j)}$. (see also Fig.~\ref{3mschema}).
The operators $\mathcal{T}_q(j)$ permute the numbers correponding to site $j$ on a SYT.
The correspondance between the $(\mathcal{T}_q(j))_{q=0...5}$ and the permutation of the symmetric group $\mathcal{S}_3$ is the following:
\begin{align}
\label{proj_3part}
\mathcal{T}_0(j)\longrightarrow& \mathcal{I}_d \\ \nonumber
\mathcal{T}_1(j)\longrightarrow& (3j-2,3j-1) \\ \nonumber
\mathcal{T}_2(j)\longrightarrow& (3j-1,3j) \\ \nonumber
\mathcal{T}_3(j)\longrightarrow& (3j-2,3j,3j-1) \\ \nonumber
\mathcal{T}_4(j)\longrightarrow& (3j-2,3j-1,3j) \\ \nonumber
\mathcal{T}_5(j)\longrightarrow& (3j-2,3j) \\ \nonumber
\end{align}
See also Fig.~ \ref{3mschema}.
Thus, for four sites with $m=3$ particles per site, in the case where $N\geq 4$, one state of the irrep $\alpha=[3,3,3,3]$ (and indeed the only one) is :
\begin{align}
\mathcal{N}^{-1}\mathcal{P}roj\,\,
 {\ensuremath \raisebox{-20pt}{$ \begin{Young}
1&4&7\cr
2&5&10\cr
3&8&11\cr
6&9&12 \cr
\end{Young} $}} 
\end{align}
which is equal to:

\begin{align}
  \frac{5}{9}\, &{\ensuremath \raisebox{-16pt}{$ \begin{Young}
1&4&7\cr
2&5&10\cr
3&8&11\cr
6&9&12 \cr
\end{Young} $}} - \frac{5\sqrt{2}}{18}\,{\ensuremath \raisebox{-16pt}{$ \begin{Young}
1&4&7\cr
2&6&10\cr
3&8&11\cr
5&9&12 \cr
\end{Young} $}} + \frac{\sqrt{5}}{3\sqrt{6}} \, {\ensuremath \raisebox{-16pt}{$ \begin{Young}
1&5&7\cr
2&6&10\cr
3&8&11\cr
4&9&12 \cr
\end{Young} $}} \nonumber \\ - \frac{5\sqrt{2}}{18} \,& {\ensuremath \raisebox{-16pt}{$ \begin{Young}
1&4&8\cr
2&5&10\cr
3&7&11\cr
6&9&12 \cr
\end{Young} $}}\,-\frac{\sqrt{5}}{6\sqrt{3}} \, {\ensuremath \raisebox{-16pt}{$ \begin{Young}
1&5&8\cr
2&6&10\cr
3&7&11\cr
4&9&12 \cr
\end{Young} $}}\,+\,\,\frac{5}{18} \, {\ensuremath \raisebox{-16pt}{$ \begin{Young}
1&4&8\cr
2&6&10\cr
3&7&11\cr
5&9&12 \cr
\end{Young} $}}\, \nonumber \\ +\frac{\sqrt{5}}{3\sqrt{6}} \, &{\ensuremath \raisebox{-16pt}{$ \begin{Young}
1&4&9\cr
2&5&10\cr
3&7&11\cr
6&8&12 \cr
\end{Young} $}}\,-\frac{\sqrt{5}}{6\sqrt{3}}  \, {\ensuremath \raisebox{-16pt}{$ \begin{Young}
1&4&9\cr
2&6&10\cr
3&7&11\cr
5&8&12 \cr
\end{Young} $}}\,\,+\,\,\,\frac{1}{6} \, {\ensuremath \raisebox{-16pt}{$ \begin{Young}
1&5&9\cr
2&6&10\cr
3&7&11\cr
4&8&12 \cr
\end{Young} $}}\,.
\end{align}
For $m\geq 4$, the construction of $\varsigma_j$ is still based on the definition of the projector shown in Eq. (\ref{projk}), but it would require
$4!=24$ coefficients.

\subsubsection{The equivalence classes in the symmetric case}
\label{symmetric_classes}
\label{eq_classes_s}
For a given  shape $\alpha$ with $N_sm$ boxes, there are $\bar{f}^{\alpha} \leq f^{\alpha} $ equivalences classes, or reprentatives. We denote them with a bar $\bar{S}_r$ (for $1\leq r\leq \bar{f}^{\alpha}$), in the same spirit as what is done in section \ref{states_as} for the antisymmetric case. They are also classified according to the last letter sequence.

Interestingly, there is an other way to define  $\bar{f}^{\alpha}$: it is the number of {\it semi-standard} Young tableaux of shape $\alpha$ and of {\it content} $1,1,..1,2,2,..2,...,N_s,N_s,..N_s$ (each number $j$ between $1$ and $N_s$ appearing $m$ times).
A {\it semi-standard} Young tableau is a tableau filled up with numbers in non-descending order from left to right in any row and in ascending order from top to bottom in any column. 
Such a number is by definition called a {\it Kostka} number (See Chapter 7 of Ref.~\onlinecite{stanley}). 
By realizing that one can pass from the antisymmetric case to the symmetric case by performing basically some {\it conjugation} of tableaux (which consists in transforming rows into columns and columns into rows), one can prove that for the same number of particle per site $m$ :
\begin{equation}
\bar{f}^{\alpha}=\tilde{f}_{\alpha^T},
\end{equation} 
where $\alpha^T$ is the transposition of the shape $\alpha$.
The decompostion of the full Hilbert space can be done exactly like in the antisymetric case, and it  leads to the following equality for the dimensions:
\begin{align}
\label{dimension_fullHS_symmetric}
\text{dim}(\bigotimes_{i=1}^{n}(\mathcal{H}\mathcal{S}_i))=(d^{+,m}_{N})^n=\sum_{\alpha} \bar{f}^{\alpha} d^{\alpha}_N.
\end{align}
Again, such an equality could be obtained as well from the Itzykson-Nauenberg rules that we review in Appendix \ref{itzykson_rules}.

\subsubsection{The basis states in the symmetric case}

We  assign to each representative SYT  $\bar{S}_r$ a specific superposition of orthogonal units $o^{\alpha}_{r1}$ that allows it to satisfy the local constraints by using the projection operator $\mathcal{P}roj=\prod\limits_{k=1}^{N_s} \mathcal{P}roj(k)$ 
where $\mathcal{P}roj(k)$ is the symmetric version of the projector defined in Eq. (\ref{projk}):
\begin{equation}
\label{projksym}
 \mathcal{P}roj(k)=\frac{1}{m!}\sum_{\sigma\in \mathcal{S}_m(k)}  \sigma.
\end{equation}
Thus, for $m=2$, and for the first site,
\begin{equation}
 \mathcal{P}roj(1)=\frac{1}{2}(\mathcal{I}_d+(1,2)),
\end{equation}
while if $m=3$:
\begin{equation}
 \mathcal{P}roj(1)=\frac{1}{6}\big\{\mathcal{I}_d+(1,2)+(1,3)+(2,3)+(1,2,3)+(1,3,2)\big\}.
\end{equation}
Then, the set of states:
\begin{align}
\nonumber
 \Big{\{} |\Psi^{\alpha}_r\rangle =\mathcal{N}^{-1}\mathcal{P}roj \times \frac{ o^{\alpha}_{r1} |\Phi^{\alpha}_1\rangle}{|| o^{\alpha}_{11} |\Phi^{\alpha}_1\rangle ||} \Big{\}}_{r=1...\bar{f}^{\alpha}} ,
\end{align}
can be proved to have the appropriate properties with the same arguments as before. In particular, the local  required symmetry is a consequence of the equality:
 \begin{align} 
P_{m(j-1)+l, m(j-1)+g} \mathcal{P}roj(j)= \mathcal{P}roj(j)\,\,\,\,\forall 1 \leq l<g\leq m.
\end{align}
Finally, by  identifying each state $ \frac{ o^{\alpha}_{p1} |\Phi^{\alpha}_1\rangle}{|| o^{\alpha}_{11} |\Phi^{\alpha}_1\rangle ||}$ with the corresponding tableau $S_p$ , one can also express the normalizing projection operator $\mathcal{N}^{-1}\mathcal{P}roj$ as an operator (for $1\leq r\leq \bar{f}^{\alpha}$) on the SYTs $
\mathcal{N}^{-1}\mathcal{P}roj \frac{ o^{\alpha}_{r1} |\Phi^{\alpha}_1\rangle}{|| o^{\alpha}_{11} |\Phi^{\alpha}_1\rangle ||}\equiv \mathcal{N}^{-1}\mathcal{P}roj \bar{S}_r =\big{(}\prod \limits_{j=1}^{N_s} \varsigma_j \big{)} \bar{S}_r ,
$
where $\varsigma_j$ is now for $m=2$:
\begin{align}
\label{coeffsymvarsigma}
\varsigma_j=\sqrt{\frac{1-\rho(j)}{2}}\mathcal{I}_d+\sqrt{\frac{1+\rho(j)}{2}}\mathcal{T}(j),
\end{align}
with the same notation as in Eq.(\ref{proj_2part}).

Note that if $2j-1$ and $2j$ are in the same line (necessarly one before the other),
then $\rho(j)=-1$ and $\varsigma_j=\mathcal{I}_d$: the permutation $P_{2j-1,2j}$ directly gives $+1$ on such a SYT.

Thus, for 6 sites with $m=2$ particles per site, in the case where $N\geq 4$, the first (out of the five) state of the irrep $\alpha=[3,3,3,3]$ is for instance:
\begin{align}
\mathcal{N}^{-1}\mathcal{P}roj\,\, {\ensuremath \raisebox{-16pt}{$ \begin{Young}
1&2&5\cr
3&4&7\cr
6&9&10\cr
8&11& 12\cr
\end{Young} $}}
\end{align}

which is equal to the normalized superposition:
\begin{align}
 \frac{3}{8}{\ensuremath \raisebox{-16pt}{$ \begin{Young}
1&2&5\cr
3&4&7\cr
6&9&10\cr
8&11& 12\cr
\end{Young} $}} +\frac{\sqrt{15}}{8}{\ensuremath \raisebox{-16pt}{$ \begin{Young}
1&2&6\cr
3&4&7\cr
5&9&10\cr
8&11& 12\cr
\end{Young} $}} +\frac{\sqrt{15}}{8}  {\ensuremath \raisebox{-16pt}{$ \begin{Young}
1&2&5\cr
3&4&8\cr
6&9&10\cr
7&11& 12\cr
\end{Young} $}} +\frac{5}{8} \, {\ensuremath \raisebox{-16pt}{$ \begin{Young}
1&2&6\cr
3&4&8\cr
5&9&10\cr
7&11& 12\cr
\end{Young} $}}.
\end{align}

For $m=3$, the operators $\varsigma_j$ are given by:
\begin{align}
\varsigma_j=\sum \limits_{q=0}^{5} \eta_q(j) \mathcal{T}_q(j),
\end{align}
where the vector of coefficients $(\eta(j))_{q=0...5}$ becomes:
\begin{align}
\label{coeffsymm}
\eta(j) = \frac{1}{\sqrt{6}} \begin{pmatrix} \sqrt{1-\rho^x(j)}\sqrt{1-\rho^y(j)}\sqrt{1-\rho^z(j)}\\ 
\sqrt{1+\rho^x(j)}\sqrt{1-\rho^y(j)}\sqrt{1-\rho^z(j)}\\
\sqrt{1-\rho^x(j)}\sqrt{1-\rho^y(j)}\sqrt{1+\rho^z(j)}\\
\sqrt{1-\rho^x(j)}\sqrt{1+\rho^y(j)}\sqrt{1+\rho^z(j)}\\
\sqrt{1+\rho^x(j)}\sqrt{1+\rho^y(j)}\sqrt{1-\rho^z(j)}\\
\sqrt{1+\rho^x(j)}\sqrt{1+\rho^y(j)}\sqrt{1+\rho^z(j)}\\
\end{pmatrix},
\end{align}
with the $\rho^a(j)$ ($a=x,y,z$) and the operators $\mathcal{T}_q(j)$ are defined as before.

Thus, for four sites with $m=3$ particles per site, in the case where $N\geq 3$, the only state of the irrep $\alpha=[3,3,3,3]$ is :
 \begin{align}
\mathcal{N}^{-1}\mathcal{P}roj {\ensuremath \raisebox{-16pt}{$ \begin{Young}
1&2&3&4\cr
5&6&7&8\cr
9&10&11&12\cr
\end{Young} $}} 
\end{align}

which is equal to:
\begin{align}
 & \frac{1}{6}\, {\ensuremath \raisebox{-16pt}{$ \begin{Young}
1&2&3&4\cr
5&6&7&8\cr
9&10&11&12\cr
\end{Young} $}} +\frac{\sqrt{5}}{6\sqrt{3}}\,{\ensuremath \raisebox{-16pt}{$ \begin{Young}
1&2&3&5\cr
4&6&7&8\cr
9&10&11&12\cr
\end{Young} $}} \nonumber \\&+\frac{\sqrt{10}}{6\sqrt{3}}\, {\ensuremath \raisebox{-16pt}{$ \begin{Young}
1&2&3&6\cr
4&5&7&8\cr
9&10&11&12\cr
\end{Young} $}} +\frac{\sqrt{5}}{6\sqrt{3}} \, {\ensuremath \raisebox{-16pt}{$ \begin{Young}
1&2&3&4\cr
5&6&7&9\cr
8&10&11&12\cr
\end{Young} $}}\,\nonumber \\&+\frac{5}{18} \, {\ensuremath \raisebox{-16pt}{$ \begin{Young}
1&2&3&5\cr
4&6&7&9\cr
8&10&11&12\cr
\end{Young} $}}\,+\,\,\frac{5\sqrt{2}}{18} \, {\ensuremath \raisebox{-16pt}{$ \begin{Young}
1&2&3&6\cr
4&5&7&9\cr
8&10&11&12\cr
\end{Young} $}}\, \nonumber \\ &+\frac{\sqrt{10}}{6\sqrt{3}} \, {\ensuremath \raisebox{-16pt}{$ \begin{Young}
1&2&3&4\cr
5&6&8&9\cr
7&10&11&12\cr
\end{Young} $}}\,+\frac{5\sqrt{2}}{18}  \, {\ensuremath \raisebox{-16pt}{$ \begin{Young}
1&2&3&5\cr
4&6&8&9\cr
7&10&11&12\cr
\end{Young} $}}\,\,\nonumber \\ &+\,\,\,\frac{5}{9} \, {\ensuremath \raisebox{-16pt}{$ \begin{Young}
1&2&3&6\cr
4&5&8&9\cr
7&10&11&12\cr
\end{Young} $}}\,.
\end{align}

\section{$SU(N)$ antiferromagnetic Heisenberg chain in the fully symmetric representation}
\label{application}
In this section, we apply this method to perform ED of the Heisenberg $SU(N)$ model on an antiferromagnetic chain in the fully symmetric irreps
with $m=2$ and $3$ particles per site.
The application to the antisymmetric case can be found in the recent paper [\onlinecite{dufourPRB2015}], where analytical predictions about the nature of the ground state (gapped or critical) due to Affleck \cite{Affleck1986a,Affleck1988} have been numerically verified by a combination of ED calculations performed along the lines of the present paper and of variational Monte-Carlo simulations.

The basic results are the energies for the $SU(N)$ symmetric chain for $m=2,3$ particles per site in the singlet subspace, and, whenever possible,
in some irreps of {\it small} quadratic Casimir.
We have employed the Lanczos algorithm whose key part is the product of the Hamiltonian (restricted to a given invariant sector) times a vector.
We have achieved this task by using a 4-step procedure.
Each basis state is represented by a SYT with proper internal constraints (see previous paragraph).
As a first step, we develop such a basis state to express it as a superposition of orthogonal units times a product state, with coefficients given by expression Eq.(\ref{coeffsymvarsigma}) (for $m=2$) and Eq.(\ref{coeffsymm}) (for $m=3$).
Then, as a second step, we apply one interaction term (corresponding to one link in the lattice) to such a superposition by employing the rules reported in the paragraph \ref{m=1}.
We first write the interaction term as a sum of permutations, like in Eq. (\ref{interaction}).
Then, each permutation is written as a product of successive transpositions whose effect on each orthogonal unit is known and described in the paragraph \ref{m=1}.
After step 2, we have a larger superposition of orthogonal units than one needs to express as a linear sum of the initial symmetric basis states.
Since the interaction term conserves the symmetry of the wave-function, one just needs to project the last superposition
using the coefficients given  by expression Eq.(\ref{coeffsymvarsigma}) (for $m=2$) and Eq.(\ref{coeffsymm}) (for $m=3$). One obtains a linear sum of symmetric states.
The final step consists in finding the ranks of those states in the ordered list of constrained SYTs (through for instance a binary search or a more sophisticated indexing function that goes beyond the scope of this paper).

In fact, since each permutation is decomposed into a product of successive transpositions, this algorithm is particularly suited for the study of chains with open boundary conditions, since it is possible to index every pair of connected sites with consecutive numbers.  \footnote{For periodic boundary conditions, we have indexed the sites in such a way that the difference between two connected sites is at most 2 by starting from 1 at some site ({\it the center}) and locating consecutive numbers alternatively to the left and to the right of this {\it center}.}
Incidentally, this also means that the computation of the exact energies is faster for open boundary conditions than for periodic boundary conditions.

\begin{figure}
\begin{center}
\includegraphics[width=260pt]{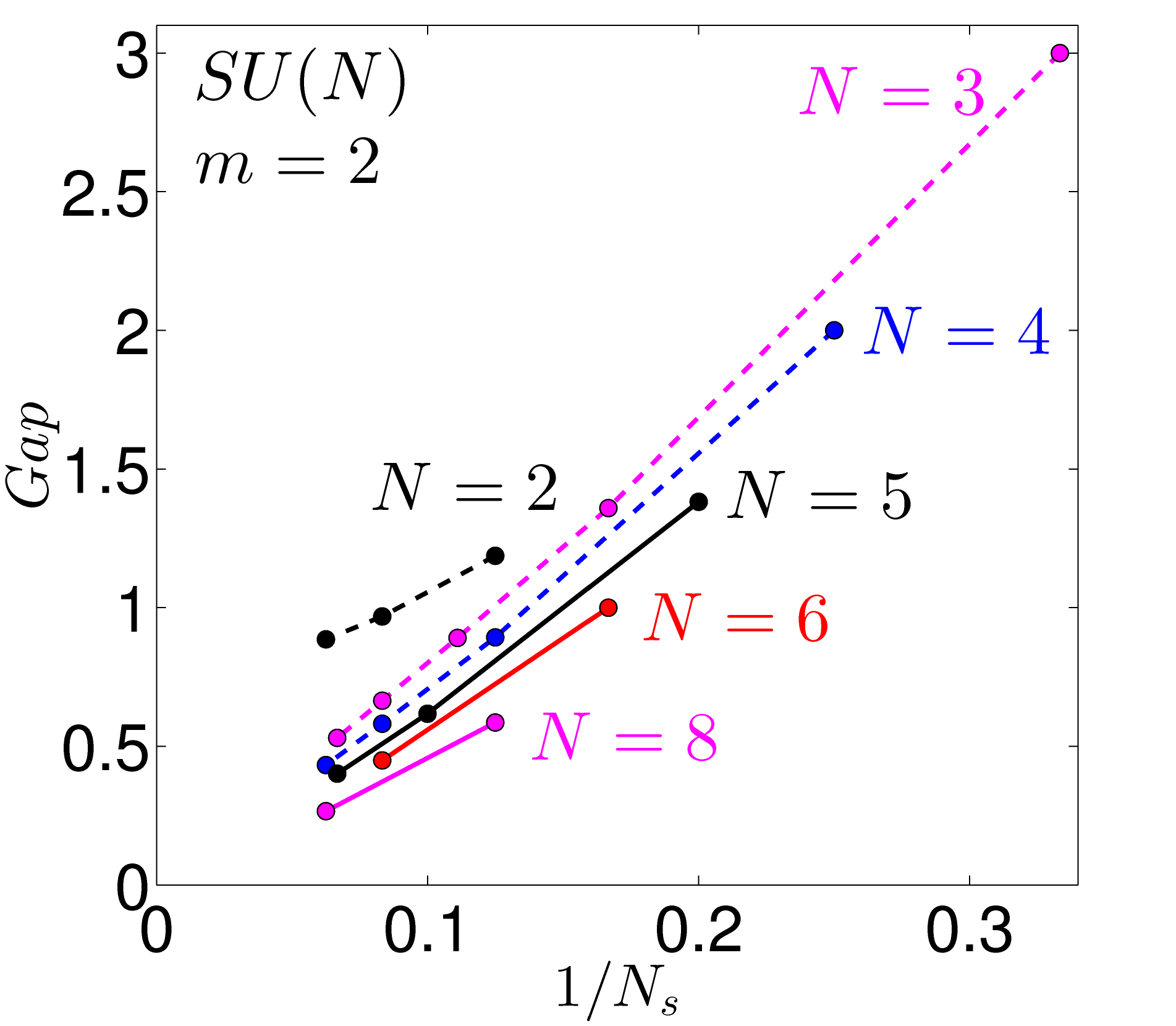}
\hskip -1cm
\caption{\label{gap_2part_sym} Gap of the $SU(N)$ antiferromagnetic Heisenberg chain with two particles per site in the symmetric irrep with periodic boundary conditions.
The gap has been determined as the energy difference between the first excited state (always located in the irrep of smaller non vanishing quadratic Casimir) and the ground state, which is a $SU(N)$ singlet 
in the systems we have considered, where the number of sites is a multiple of $N$.
For $SU(2)$ (the spin-1 chain), one can infer the presence of the Haldane gap from 
the results obtained for relatively small chains
(up to $N_s=16$), whereas, for $N\geq3$, the results for similar sizes are consistent with a gapless spectrum in the thermodynamic limit.
}
\vspace{-0.5cm}
\end{center}
\end{figure}

For $m=2$, we list in Table \ref{arraysinglets} the ground state energies per site for periodic boundary conditions ($ \mathcal{E}_{GS}^P(N_s)$), as well as for open boundary conditions ($ \mathcal{E}_{GS}^O(N_s)$) since it can be useful for benchmarking future DMRG studies. When the number of sites $N_s$ is a multiple of the number of colors $N$, $N_s=pN$, the ground state is always a singlet (for antiferromagnetic couplings),
so that the minimal energy is obtained by diagonalizing the Hamiltonian in the $SU(N)$ singlet sector, i.e the sector corresponding to the shape 
$\alpha=[q,q,...,q]$, where $q=N_s m/N=pm$.
We also provide in Table \ref{arraysinglets} the corresponding dimensions $\bar{f}^{[q,...,q]}$ that give the size of the matrices we diagonalized.

\begin{table}[h]
	\centering
	\begin{tabular}{|c|c|c|c|c|}
		\hline
		 m=2 & $N_s$ &$\bar{f}^{[q,...,q]}$ &$\mathcal{E}_{GS}^P(N_s)$ & $ \mathcal{E}_{GS}^O(N_s) $\\\hline
		SU(3) &15&6879236&-1.448589&-1.397889\\
		SU(3) &18&767746656& &-1.402602\\
		 SU(4) &16&190720530&-1.687431&-1.619271\\
	           SU(5) &15 &25468729&-1.804955&-1.716275\\
		 SU(6)  &12 &16071& -1.88243593&-1.752105\\
		 SU(8)& 16& 3607890& -1.932087&-1.827798\\
		SU(10)& 20& 1135871490& &-1.869078\\
		\hline
	\end{tabular}
	\caption{
		Dimension of the singlet sector $\bar{f}^{[q,...,q]}$, ground state energy per site with periodic boundary conditions $\mathcal{E}_{GS}^P(N_s) $, and ground state energy per site with open boundary conditions $\mathcal{E}_{GS}^O(N_s) $ for $SU(N)$ Heisenberg chains with two particles per site in the
symmetric irrep and $N_s$ sites. The calculatation is more difficult for periodic boundary conditions because the matrix is less sparse, and it has not been done for $SU(10)$ with $20$ sites and for $SU(3)$ with $18$ sites since it would require a parallel architecture with several hundreds of nodes. 
}
	\label{arraysinglets}
\end{table}

\subsection{Gap}

We have applied this algorithm to determine the gap of the Heisenberg $SU(N)$ symmetric chain.
With both periodic ($\mathcal{E}_{ex}^P(N_s)$) and open ($\mathcal{E}_{ex}^0(N_s)$) boundary conditions, as long as the number of sites $N_s$ is a multiple of $N$, the first excited state belongs to the {\it adjoint} irrep $\alpha=[q+1,q,..,q-1]$ with the smallest non vanishing quadratic Casimir $C_2$.
Its dimension $\bar{f}^{[q+1,q,..,q-1]} $ is always much larger than that of the singlet. 
For instance, for $SU(10)$ and $20$ sites, $\bar{f}^{[5,4,4,4,....,4,3]}\approx 40 \times 10^9 $, a size we could not handle with the computers at our disposal,
whereas the dimension of the singlet sector is $\approx 1.3 \times 10^9 $.
We have gathered some values of $\mathcal{E}_{ex}^P(N_s)$ and $\mathcal{E}_{ex}^0(N_s)$ in Table \ref{arrayexcited}.

\begin{table}[h]
	\centering
	\begin{tabular}{|c|c|c|c|c|}
		\hline
		 m=2 & $N_s$ & $\bar{f}^{[q+1,q,..,q-1]}$ &  $\mathcal{E}_{ex}^P(N_s)$  &  $\mathcal{E}_{ex}^O(N_s)$ \\\hline
		 SU(3) &15&44994040&-1.413231&-1.381094\\
\hline SU(4) &16&2077175100&-1.6604222&-1.607230\\
\hline SU(5) &15 &377182806&-1.778192&-1.705462\\
\hline  SU(6)  &12 &272712&-1.84503164&-1.738918\\
\hline   SU(8)& 16&93683590&-1.915457&-1.822196 \\
		\hline
	\end{tabular}
	\caption{Dimension of the representation of smallest non-vanishing Casimir $\bar{f}^{[q+1,q,..,q-1]}$  (corresponding to the {\it triplet} sector for $SU(2)$), and first excited energies per site in that sector for $SU(N)$ Heisenberg chains with two particles per site in the
symmetric irrep and $N_s$ sites for periodic boundary conditions ($\mathcal{E}_{ex}^P(N_s)$ ) and open boundary conditions ($\mathcal{E}_{ex}^O(N_s)$). 	
}
	\label{arrayexcited}
\end{table}

For two particles per site, we have plotted the corresponding gaps in Fig. \ref{gap_2part_sym}, in which the $SU(2)$ case has been added for comparison.
The $SU(2)$ case corresponds to the spin-1 chain known to exhibit the Haldane gap,
which can be clearly inferred from the results obtained even for relatively small chains ($N_s\leq16$).
Note the factor $2$ between our interpolated value (around 0.8) and the DMRG value taken from \cite{whiteprb1993} ($\Delta\approx 0.410$), which comes from the two different ways to write the interaction between $2$ sites, either in terms of spin operators or in terms of permutation  operators. Indeed, if we use the notations of the previous section, the spin $1$ interaction between sites $1$ and $2$ is related to the permutations through the identity
$\vec{S}_1.\vec{S}_2=\frac{1}{2}\{P_{1,3}+P_{1,4}+P_{2,3}+P_{2,4}\}-1$.

In all cases except $SU(2)$, the data are consistent with a vanishing gap in the thermodynamic limit, hence with a gapless spectrum. So the difference between
$1$ and $2$ particles per site predicted by Haldane for $SU(2)$ does not seem to carry over to larger values of $N$.

For three particles per site, the sizes we can reach with our algorithm are smaller ($N_s\leq12$). 
The ground state energies are listed in Table \ref{arraysinglets,3}.

\begin{table}[h]
	\centering
	\begin{tabular}{|c|c|c|c|c|}
		\hline
		m=3 &$N_s$ &$\bar{f}^{[q,...,q]}$ &  $\mathcal{E}_{GS}^P(N_s)$ &  $\mathcal{E}_{GS}^O(N_s)$\\
\hline SU(3) &12&3463075&-2.218913&-2.106611\\
\hline SU(4) &12&10260228&-2.574628&-2.421300\\
\hline  SU(6)  &12 &1113860&-2.832493&-2.634385\\
		\hline
	\end{tabular}
	\caption{Dimension of the singlet sector $\bar{f}^{[q,...,q]}$, ground state energy per site with periodic boundary conditions $\mathcal{E}_{GS}^P(N_s) $, and ground state energy per site with open boundary conditions $\mathcal{E}_{GS}^O(N_s) $ for $SU(N)$ Heisenberg chains with three particles per site in the
symmetric irrep and $N_s$ sites
	}
	\label{arraysinglets,3}
\end{table}

\begin{figure}
\begin{center}
\includegraphics[width=260pt]{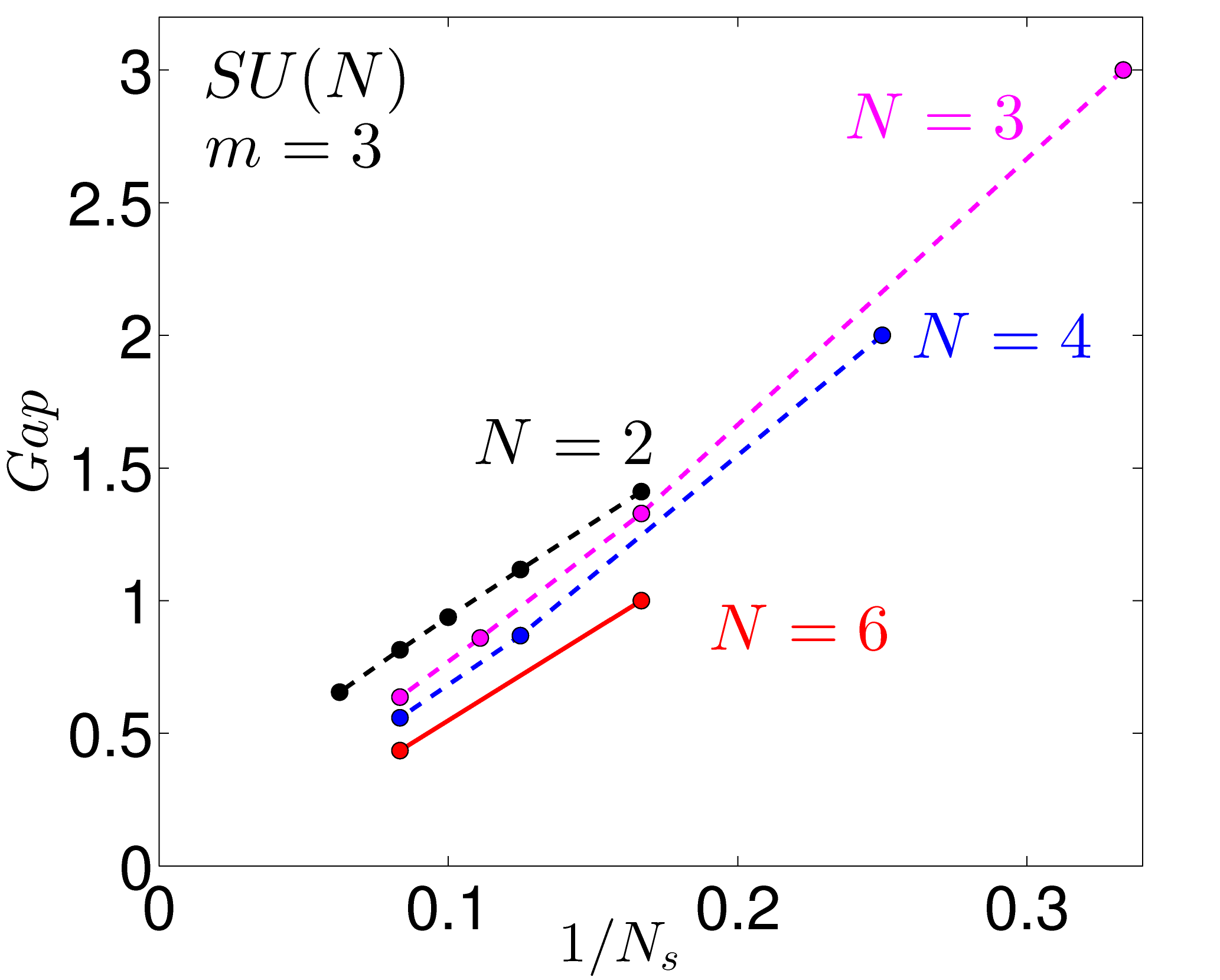}
\vskip -0.2cm
\caption{\label{gap_3part_sym} Gap of the $SU(N)$ antiferromagnetic Heisenberg chain with three particles per site in the symmetric irrep with periodic boundary conditions.
The gap has been determined as the energy difference between the first excited state (always located in the irrep of smaller non vanishing quadratic Casimir) and the ground state, which is a $SU(N)$ singlet 
in the systems we have considered, where the number of sites is a multiple of $N$.
In the case $N=2$ (spin-3/2 chain), which is known to be gapless, the curve is slightly convex, consistent with a vanishing gap in the thermodynamic limit. For $N=3,4,6$, the concavity is opposite and the curves seem to converge towards zero; however the maximal sizes reached in the simulations are quite small ($N_s\leq 12$).
}
\vspace{-0.5cm}
\end{center}
\end{figure}

Quite surprisingly, the results for the gap are also consistent with a vanishing gap in the thermodynamic limit  for $SU(3)$, $SU(4)$ and $SU(6)$, as shown in Fig. \ref{gap_3part_sym}. For $SU(3)$, these results are in contradiction with the DMRG results reported in Ref.~\onlinecite{rachel2009}. We will come back to
this difference in the discussion section below.

\subsection{Central charge}

When a 1D quantum system is critical, it is in general possible to identify the universality class to which the low energy theory belongs in terms
of the underlying conformal field theory (CFT).
In the case of Heisenberg $SU(N)$ models, the relevant CFTs are the $SU(N)$ Wess-Zumino-Witten (WZW) models with topological integer coupling coefficient $k$ \cite{affleck1986}. The corresponding algebra of such a theory is $SU(N)_k$, with a central charge $c$ given by:
\begin{equation}
\label{centralSUk}
c=k\frac{N^2-1}{N+k}.
\end{equation}

\begin{figure}
\begin{center}
\includegraphics[width=250pt]{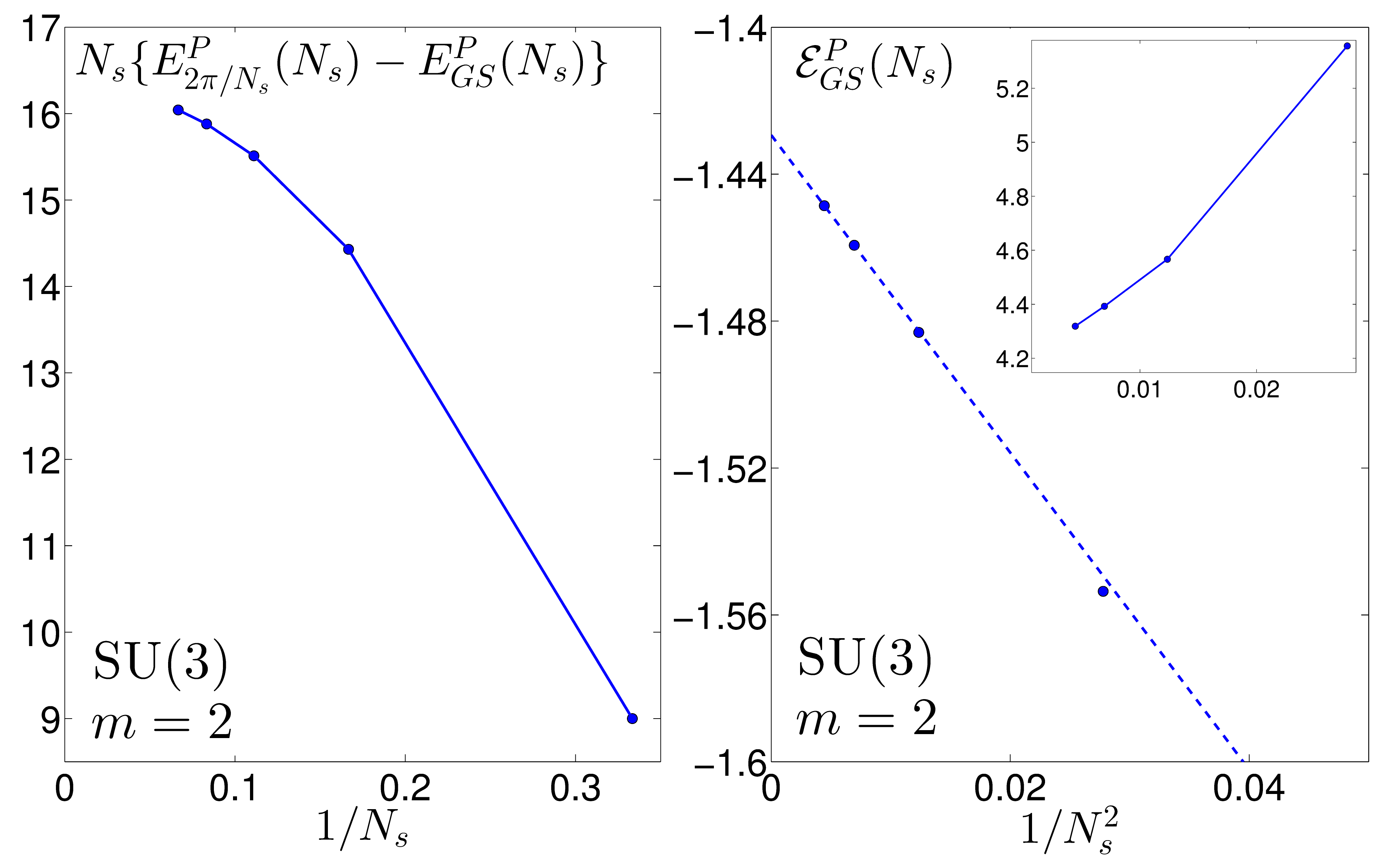}
\vskip -0.2cm
\caption{\label{fitting} 
Examples of finite size results that were used to extract the central charge for $SU(3)$ and $m=2$. Left: Excitation energy of the first excited
state with momentum $k=2\pi/N_s$ times the number of sites as a function of  $1/N_s^2$. In a critical system, this is expected to
tend to $2\pi v$ in the thermodynamic limit, where $v$ is the sound velocity. Right: 
Ground state energy per site $\mathcal{E}_{GS}^P(N_s)$ as a function of $1/N_s^2$.
In a critical system, it is expected to converge to $\mathcal{E}_{GS}^P(\infty)$ linearly with $1/N_s^2$, with a slope equal to $2\pi c v/12$, where $c$ is the central charge. 
}
\vspace{-0.5cm}
\end{center}
\end{figure}

The central charge can be extracted from the exact diagonalization results in two steps. First of all, one can extract the product of the central charges $c$ with
the sound velocity $v$ from the dependence of the ground state energy per site $\mathcal{E}_{GS}^P(N_s)$ with the number of sites $N_s$\cite{affleck1986,bloteprl56,moreo}:
\begin{equation}
\label{energy_charge}
\mathcal{E}_{GS}^P(N_s)=\mathcal{E}_{GS}^P(\infty)-\frac{2\pi c v}{12 N_s^2} + o(1/N_s^2),
\end{equation}
where $\mathcal{E}_{GS}^P(\infty)$ is the ground state energy per site in the thermodynamic limit.
The case $SU(3)$ with $m=2$ is shown as an example in the right panel of Fig.~\ref{fitting}. The scaling as $1/N_s^2$ is already quite accurate for the largest
available sizes. 

Secondly, one can extract the sound velocity $v$ from the energy of the first excited state of momentum $k=2\pi/N_s$ and non zero quadratic
Casimir\cite{cardy}:
\begin{equation}
\label{velocity}
E_{2\pi/N_s}^P(N_s)-E_{GS}^P(N_s)= \frac{2\pi v}{N_s} +o(1/N_s).
\end{equation}
where $E_{2\pi/N_s}^P(N_s)$ and $E_{GS}^P(N_s)$ are total energies. 
To check the momentum of the excited state, which is not available right away in our approach since we do not use spatial symmetries, we 
had to extract the ground state wave function, and to apply directly the translation operator. This is tedious but straightforward because the translation
operator can be written in terms of permutations. It turns out that in all the gapless cases investigated ($m=2,3$ for $N>2$), and as soon as $N_s=pN$ with $p>1$, the first excited state in the adjoint irrep has a momentum $k=2\pi/N$, and it is actually only the second excited state which has
the momentum $k=2\pi/N_s$. Examples of the resulting finite-size estimates of the velocity $v$ are given in the left panel of Fig. \ref{fitting} for $N=3$, $m=2$. 

To avoid uncertainties due to extrapolations in extracting the central charge $c$, we have used for the velocity $v$ the value for the largest available size, and 
for the product $cv$ the slope deduced from the values of the ground state energy for the two largest sizes. 
The central charges extracted in this way can be expected to be slightly overestimated since the product $cv$ decreases with the size while the velocity $v$ increases with the size.
The corresponding estimates for the central charge are listed in Table \ref{centralchargesm2} for $m=2$ and in Table \ref{centralchargesm3} for $m=3$, together with the theoretical values for $SU(N)_1$ and $SU(N)_m$. Quite remarkably, in all cases, the results $m$ particles per site for are in good agreement with $SU(N)_m$ (and slightly above, as expected), and very far from $SU(N)_1$. 

This result is quite surprising since, according to field theory, the $SU(N)_{k>1}$ WZW models have at least one relevant operator allowed by symmetry\cite{Affleck1988,Lecheminant2015b},
implying that one should adjust at least one parameter to sit at such a critical point, as in the case of integrable models\cite{andrei1984}.

\begin{table}[h]
	\centering
	\begin{tabular}{|c|c|c|c|c|c|c|c|}
		\hline
		&&SU(3)&SU(4)&SU(5)&SU(6)&SU(7)&SU(8) \\
		\hline
		m=2&c&3.23&5.16&7.16&9.77&11.65&13.56 \\
		\hline
	k=2&$\frac{N^2-1}{N+2}$ &3.2&5&6.86&8.75&10.67&12.60\\
\hline
	k=1&$\frac{N^2-1}{N+1}$ &2&3&4&5&6&7\\
\hline
	\end{tabular}
	\caption{Finite-size estimates of the central charge $c$ (see main text and Fig. ~\ref{fitting}) for $SU(N)$ Heisenberg chains with two particles per site in the symmetric irrp ($m=2$), compared to the predictions for the
	  $SU(N)_2$ and $SU(N)_1$ WZW universality classes.
	  }
	\label{centralchargesm2}
\end{table}

\begin{table}[h]
	\centering
	\begin{tabular}{|c|c|c|c|c|c|}
		\hline
		&&SU(3)&SU(4)&SU(6)\\
		\hline
		m=3&c&4.09&7.49&13.21 \\
		\hline
	k=3&$\frac{N^2-1}{N+3}$ &4&6.43&11.67\\
\hline
	k=1&$\frac{N^2-1}{N+1}$ &2&3&5\\
\hline
	\end{tabular}
	\caption{Finite-size estimates of the central charge $c$ (see main text and Fig. ~\ref{fitting}) for $SU(N)$ Heisenberg chains with three particles per site in the symmetric irrp ($m=3$), compared to the predictions for the
	  $SU(N)_3$ and $SU(N)_1$ WZW universality classes.
	  }
	\label{centralchargesm3}
\end{table}

\subsection{Scaling dimension}

\begin{figure}
\begin{center}
\includegraphics[width=255pt]{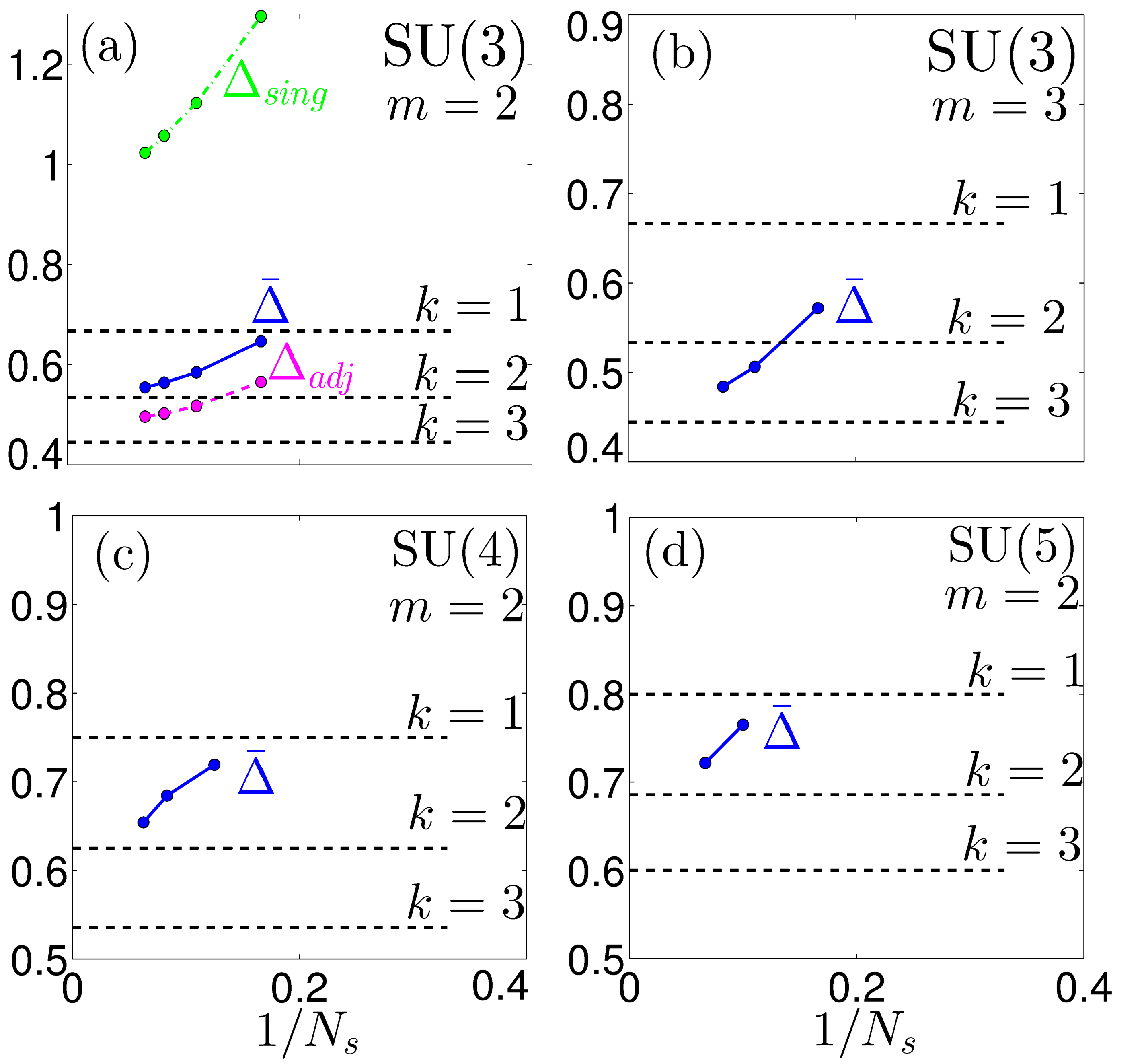}
\vskip -0.2cm
\caption{\label{critical_dimension} 
Finite-size estimates of the critical dimension $\bar{\Delta}=\frac{(N^2-1)\Delta_{adj}+\Delta_{sing}}{N^2}$ averaged between the singlet estimate
$\Delta_{sing}$ and the adjoint estimate $\Delta_{adj}$ to suppress the main logarithmic corrections in a few representative case. Dashed lines: predictions
for the WZW $SU(N)$ level $k$ theory. For $SU(3)$ and
$m=2$, the singlet and adjoint estimates are also included for comparison. In all cases, the results are consistent with the prediction
$\Delta=\frac{N^2-1}{N(N+k)}$ for the WZW $SU(N)$ level $k=m$, and clearly below that of the WZW $SU(N)$ level $k=1$.
}
\vspace{-0.5cm}
\end{center}
\end{figure}

To further check this identification, it is possible to extract additional information from the spectra, namely the scaling dimension $\Delta$ of the primary fields.
Indeed, the other low-lying excited energies should satisfy some scaling relations analogous to Eq. (\ref{velocity}) \cite{moreo}:
\begin{equation}
\label{critsing}
E_{exc, sing}^P(N_s)-E_{GS}^P(N_s)= \frac{2\pi v \Delta}{N_s}+o(1/N_s),
\end{equation}
where $E_{exc, sing}^P(N_s)$ is the first excited energy in the singlet subspace, and 
\begin{equation}
\label{critadj}
E_{adj}^P(N_s)-E_{GS}^P(N_s)= \frac{2\pi v \Delta}{N_s}+o(1/N_s),
\end{equation}
where $E_{adj}^P(N_s)$ is the lowest energy in the adjoint subspace (irrep $[q+1,q,..,q-1]$).
Combined with Eq. (\ref{velocity}), these scaling relations allow one to obtain finite-size estimates of the scaling dimension as ratios of excitation energies according to:
\begin{equation}
\label{critsing}
\Delta_{sing}=\frac{E_{exc, sing}^P(N_s)-E_{GS}^P(N_s)}{E_{2\pi/N_s}^P(N_s)-E_{GS}^P(N_s)},
\end{equation}
and
\begin{equation}
\label{critsing}
\Delta_{adj}=\frac{E_{adj}^P(N_s)-E_{GS}^P(N_s)}{E_{2\pi/N_s}^P(N_s)-E_{GS}^P(N_s)}.
\end{equation} 
In the thermodynamic limit, both estimates $\Delta_{sing}$ and $\Delta_{adj}$ should converge to the scaling dimension of the primary field of the $SU(N)_k$ theory, $\Delta$, which
is given by:
\begin{equation}
\Delta=\frac{N^2-1}{N(N+k)}.
\end{equation} 
However, the scaling to the thermodynamic limit is in general very slow because of logarithmic corrections. One way to get around this difficulty, pioneered
for $SU(2)$ by Ziman and Schulz\cite{ziman1987}, consists in getting rid of the main logarithmic corrections by considering the linear combination 
$\frac{(N^2-1)\Delta_{adj}+\Delta_{sing}}{N^2}$ as an estimate of $\Delta$. 
The resulting estimates of the critical dimension are compared to the theoretical predictions for a few specific cases in Fig.~\ref{critical_dimension}.
As expected, the results for $\Delta_{sing}$ and $\Delta_{adj}$ can be very different (see Fig.~\ref{critical_dimension}a), but once the appropriate
linear combination is considered, the results are again consistent with the universality class $SU(N)_m$ for $m$ particles per site.

\subsection{Discussion}

All the ED numerical evidence collected on systems with $N>2$ and $m=2$ or $3$ is consistent with a gapless spectrum and a critical behavior
in the $SU(N)_m$ WZW universality class. These results are in apparent contradiction with previous analytical and numerical results. 

Quite generally, renormalization group (RG) arguments seem to exclude $SU(N)_m$ WZW as a generic critical theory for $N>2$ and $m>1$ because there
is a relevant operator allowed by symmetry that should either open a gap or drive generic systems away from this critical point towards 
the stable $SU(N)_1$ WZW critical point under the RG flow\cite{Affleck1988,Lecheminant2015b}. What our results suggest is that, for intermediate energies and length scales, the physics
is indeed governed by the $SU(N)_m$ WZW critical theory, and that the unstable nature of this critical point will only show up as a cross-over 
at length scales larger than the size of the biggest clusters we have studied. This is reminiscent of the spin-3/2 chain studied by Ziman and Schulz\cite{ziman1987} and by Moreo\cite{moreo}, in which finite-size estimates of the central charge and of the scaling dimension were changing 
significantly with the system size. However, in
the case of the spin-3/2 chain, results consistent with $SU(2)_1$ WZW were already obtained for systems with 12 sites, whereas in our case the results
are still fully consistent with $SU(N)_m$ WZW for systems with up to 18 sites for $m=2$. So, if there is a cross-over, it has to take place for rather
large length scales.

The DMRG results for 48 sites reported in Ref.~\onlinecite{rachel2009} for $SU(3)$ with $m=2$ and $m=3$ are also at variance with the conclusions 
drawn from ED of small systems. For $m=2$, a fit of the entanglement entropy with the Calabrese-Cardy formula has led to the estimate $c=2.48$ for the central charge, in agreement with the $SU(3)_1$ WZW universality class with central charge $c=2$ because of logarithmic corrections according to the authors of Ref.~\onlinecite{rachel2009}, while for $m=3$, the saturation of the entanglement entropy has been taken as an evidence that the system is gapped. 
At first sight, these results are consistent with the cross-over scenario: for $m=2$, the system would be in the middle of the cross-over between $SU(3)_2$
and $SU(3)_1$ for 48 sites, while for $m=3$, the gap is already well developed for this size. 

There are however a number of puzzling aspects. For $m=3$,
the DMRG results of the entanglement entropy have already saturated after 10 sites, which suggests that the correlation length is smaller than 10. This
is inconsistent with the results of Fig.~\ref{gap_3part_sym}, which show no sign of a gap for 12 sites. For comparison, the presence of a gap for the
Haldane chain is already visible on smaller systems. 
In a similar spirit, for $m=2$, the central charge is already much smaller than $c=3.2$, the theoretical value for $SU(3)_2$, for 48 sites, which suggests
that the cross-over has already started long before. This is not obviously consistent with the results of Fig.~\ref{gap_2part_sym}, where there is no 
sign of any significant curvature for $N=3$ up to 15 sites. So we think that the presence of a cross-over and its characteristic length-scale require
further investigation.

\section{Conclusion}
We have developed a method to perform exact diagonalizations of Heisenberg $SU(N)$ models with $m$ particles per site 
in the fully symmetric and antisymmetric irreps directly in the symmetry sectors of the global $SU(N)$ symmetry of the problem,
thereby allowing one to reach larger cluster sizes than with the traditional approach. 
The central result is that the relevant orthonormal basis is in one-to-one correspondance with some subset of standard Young tableaus.
We have provided the details of the rules to select this subset, an efficient way to computationally generate them, and the rules to write $SU(N)$
interaction in this basis. 

We have applied this formalism to the investigation of the symmetric Heisenberg $SU(N)$ chain with two and three particles.
For both $m=2$ and $m=3$, the finite-size results are consistent with a gapless spectrum for any value of $N\geq3$, and 
finite-size estimates of the central charge and of the scaling dimension of the primary field suggest that the physics is governed
by the $SU(N)_m$ WZW universality class. In view of previous results based on renormalization group arguments and on DMRG 
simulations, we suspect that a crossover towards a gapped state or towards an $SU(N)_1$ theory might take place
when increasing the system size, a possibility that requires further investigation however.

We have also provided some results for systems with open boundary conditions that might be useful fo benchmark DMRG studies 
that take advantage of the $SU(N)$ symmetry, an issue of great current interest \cite{weichselbaum2012,mcculloch2007,mcculloch2002,gvidal2010}.

Finally, we also plan to extend the method to other irreps in order to study $SU(N)$ AKLT spin chains\cite{rachel2007,morimoto2014} that can lead to a rich variety of symmetry protected topological phases (\cite{nonne2013,quella2012,quella2013}), or their generalisation in $2D$, the $SU(N)$ simplex phases\cite{arovas2008}.

{\it Acknowledgements:}  
We thank Ian Affleck, Sylvain Capponi, Philippe Lecheminant, Thomas Quella, and Hong-Yu Yang for useful discussions and advices. 
This work has been supported by the Swiss National Science Foundation. 

\section{Appendix}
\subsection{Fermionic and bosonic representation of the Hamiltonian in the antisymmetric and symmetric cases}
\label{fermionic}
For the totally symmetric (resp. antisymmetric) irrep, the $SU(N)$ generators can be expressed as (we drop the upper index which stands for the label of the site):

\begin{equation*}
	\hat{S}_{\alpha\beta} = \hat{f}_{\alpha}^{\dag}\hat{f}_{\beta} - \dfrac{m}{N}\delta_{\alpha\beta}
\end{equation*}
where $m$ is the number of particles per site, 
where  $f^{\dag}_{\alpha}$ and $f_{\alpha}$ are the creation and annihilation
operators for a boson (resp. fermion)  of color $\alpha=1,...,N$.
Thus, the interaction term between sites $i$ and $j$ becomes:
\begin{align}
\label{interaction_ferm_ij}
H_{(i,j)}= \sum_{\mu, \nu}f^{\dag}_{\mathbf{i}\mu} f_{\mathbf{i}\nu} f^{\dag}_{\mathbf{j}\nu} f_{\mathbf{j}\mu} ,
\end{align}
where we have dropped the constant $-m^2/N$.
The local basis is defined by filling up each site with $m$ particles. Then, it is easy to show that it is possible to express the last Hamiltonian 
as a sum of $m^2$  permutations. Indeed,
assigning number $m(j-1)+l$ to the $l^{th}$ particle of site $j$, for $l=1...m$, one can prove that it is equivalent to Eq.~(\ref{interaction}).

\subsection{Proof that the projection gives zero for non relevant SYTs. }
\label{proof_1}
Let us discuss the antisymmetric case to fix the ideas (the arguments are the same for the symmetric case).
We are going to prove that if two different numbers $q$ and $q'$ of the set $\{m(k-1)+1,\,\,m(k-1)+2\, ...,\,mk\}$ belong to the same line in an SYT $S_r$, then $Proj(k) S_r=0.$
First of all, $q$ and $q'$ can be chosen to be in adjacent boxes of the same line (necessarily, there is no number between $q$ and $q'$ that would correspond to a site $k'>k$ since $S_r$ is a SYT).
Then, we first consider the simple case where $q'=q+1$.
Since $Proj(k)$ can be factorized on the right by $(I_d-(q,q+1))$ \cite{rutherford}, one directly has $Proj(k)S_r\propto (I_d-(q,q+1)) S_r=0$ thanks to the rules controling the effect of the permutation $(q,q+1)$ on a tableau (cf section \ref{m=1}).
Let us now focus on the case where $q'>q+1$ (still in two adjacent boxes).
The first thing to notice is that there is a transformation $\sigma_{rp}(k)$  that allows one to pass from $S_r$ to a tableau $S_p$, where all the numbers but the ones of site $k$ would be at the same location as in $S_r$, and with the nunmbers of site $k$ permuted in such a way that $q+1$ would be at the location of $q'$.
Such a  $\sigma_{rp}(k)$ is very much like the operator $\mathcal{T}(k)$ (in Eq. (\ref{proj_2part}) for $m=2$) or like   $\mathcal{T}_q(k)$ ($q=0...5$ in Eq. (\ref{proj_3part}) for $m=3$): it exchanges numbers on a SYT, such that:
\begin{equation}
\sigma_{rp}(k) S_p \rightarrow S_r.
\end{equation}
$\sigma_{rp}(k)$ corresponds to a linear combination of permutation  $\eta_{rp}(k)$ (involving permutation between numbers of particles of site $k$) such that:
\begin{equation}
o^{\alpha}_{r1}=\eta_{rp}(k)o^{\alpha}_{p1}.
\end{equation}
Now, the important thing to notice is that $Proj(k)$ is proportionnal to the orthogonal units $o^{[1 1 ...1]}_{11}$ , where $[1 1 ...1]$ is the $m$ boxes fully antisymmetric irrep, (with number $1,2,...,m$ replaced by numbers $m(k-1)+1,\,\,m(k-1)+2\, ...,\,mk$).
For this one column shape, there is just one SYT and so just one orthogonal unit  $o^{[1 1 ...1]}_{11}$.
Then, the orthonormal properties in Eq. (\ref{relation_orthonormal}) imply that:
\begin{align}
&Proj(k) \, o^{\alpha}_{r1}= Proj(k) \eta_{rp}(k) o^{\alpha}_{p1}\propto Proj(k) \, o^{\alpha}_{p1},
\end{align}
which is zero thanks to the case considered just before.

\subsection{Review of the  Itzykson-Nauenberg rules}
\label{itzykson_rules}
\begin{figure}
\begin{center}
\includegraphics[width=250pt]{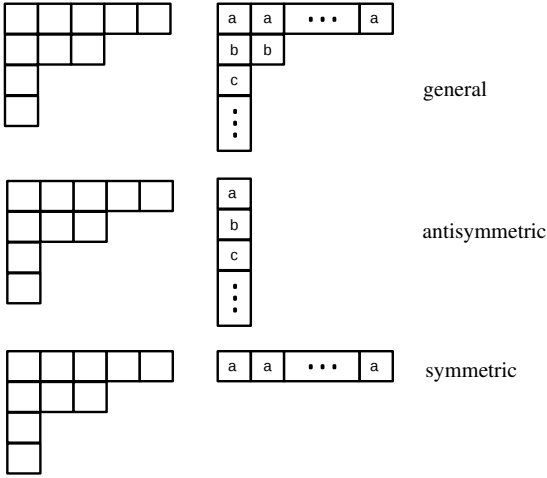}
\caption{\label{tensor_product} Itzykson-Nauenberg recipe (see text for details) to perform the tensorial producf of the $SU(N)$ irrep on the left times the $SU(N)$ irrep on the right in the general case ( top panel), antisymmetric case (middle panel) and symmetric case (bottom panel).}
\end{center}
\end{figure}

Let us make the tensorial product of two arbitrary representations, like those shown in the top panel of Fig. \ref{tensor_product}.
Choose one of those as the "trunk" (the left one in the figure  \ref{tensor_product}), and label the boxes in the first row of the second tableau with "a",
the boxes in the second row with "b", etc...
Add one box labelled "a" on the trunk in all possible ways such that it remains a tableau (length of rows in non increasing order from top to bottom).
Then, add a second box labelled "a" (if any) requiring that the resultant object is still a tableau, etc...
When the boxes labelled "a" are exhausted in the second tableau, add the boxes  labelled "b", etc...
In this process, satisfy the following rules:\\
(i)Keep only the tableaux with no more than N rows.\\
(ii)Never let two boxes with the same label stand in the same column.\\
(iii)Reading from right to left and top to bottom a resulting tableau, collect the labels of the boxes.\\
One should always find a number of "a"s greater or equal to the number of "b"s, which itself should be greater or equal to the number of "c"s, and so on.\\
(iv)Tableaux with the same attached labels at the  same place should be counted as one. That is to say, two identical representations in the resulting tensorial product of the two shapes
should differ by the disposition of the letters.\\
For one particle per site, one can perform $N_s$ times the tensorial product with the fundamental irrep (one box) to see how many times each collective representations appear (their {\it multiplicity})
in the full Hilbert space.
Numbering the boxes by the step at which the box is added to the current shape (corresponding also to the number of the particle added), the Itzykson-Nauenberg rules clearly involve that the number of times each $N_s$- boxes Young tableau $\alpha$
appear in the full Hilbert space is $f^{\alpha}$, i.e the number of SYTs of shape $\alpha$.
If now we consider several particles per site, assigning number $m(j-1)+l$ to the $l^{th}$ particle of site $j$, for $l=1...m$, the Itzykson-Nauenberg rules involve some selection over all the SYTs of $N_s \times m$
boxes of shape $\alpha$ appearing in the full Hilbert space.
The selected SYTs must respect internal constraints that depend on the nature of the local $SU(N)$ symmetry (either fully antisymmetric or symmetric).

For the antisymmetric  irrep with $m-$boxes at each site, one has to perform tensorial product of the kind shown in the middle panel of Fig.~\ref{tensor_product}. If one replaces "a", "b", "c", and so on by the particle numbers of each site, that is to say the numbers $m(k-1)+1,\,\,m(k-1)+2\, ...,\,mk$ (where $1\leq k\leq N_s$ is the index of the site), the rule (iii) above implies that those numbers should be located in rows $y(m(k-1)+1)<y(m(k-1)+2)<...<y(mk)$, 
where $y(q)$ is the row (between $1$ and $N$) where the number $1\leq q\leq mN_s$ is located in the considered SYT. 
This rule involves a precise selection over all the $ f^{\alpha}$ SYTs of shape $\alpha$, which is nothing but the selection of the $\tilde{f}_{\alpha}$ representatives in the antisymmetric case (see section \ref{eq_classes_as}).
For the symmetric $m-$boxes irrep at each site, the bottom panel of figure \ref{tensor_product} shows the kind of tensorial product
to be done. 
In fact, if one replaces the $m$ letters 'a' of the row irrep of site $k$ by  the particle  numbers $m(k-1)+1,\,\,m(k-1)+2\, ...,\,mk$ (where $1\leq k\leq {N_s}$) , the  Itzykson-Nauenberg rule (ii) implies that those numbers must be located in columns $c(m(k-1)+1), c(m(k-1)+2),....,c(mk)$, which are all different from each other,  $c(q)$ being defined in general as the column (between $1$ and $\alpha_1$) where the number $1\leq q\leq mN_s$ is located in the considered SYT. 
Secondly, the  Itzykson-Nauenberg rule (iv) implies that for any acceptable set of locations for the numbers  $m(k-1)+1,\,\,m(k-1)+2\, ...,\,mk$ (where $1\leq k\leq {N_s}$), only one configuration should be kept. For instance, one can keep the one where  the numbers   $m(k-1)+1,\,\,m(k-1)+2\, ...,\,mk$ are located  in ordered rows such that $y(m(k-1)+1)\leq y(m(k-1)+2)\leq...\leq y(mk)$. So it is also equivalent to the selection of the representatives in the symmetric case  (see section \ref{eq_classes_s}).
Thus, Eq. (\ref{dimension_fullHS_antisymmetric}) and (\ref{dimension_fullHS_symmetric}) are direct consequences of the  Itzykson-Nauenberg rules and can be obtained regardless of the projection procedure.

\subsection{Proof that each class leads to a unique linear combination of permutation (up to a constant) after projection. }
\label{proof_2}
Let us discuss the antisymmetric case to fix the ideas (the arguments are the same for the symmetric case).
If two indices $r$ and $p$ label SYTs belonging to the same class $S_r$ and $S_p$, it means that there is a permutation $\sigma_{rp}$  that allows one to pass from $S_r$ to $S_p$, like in section \ref{proof_1}.
By definition of a class, it means that $\sigma_{rp}$ can be factorized into a product of $N_s$ permutations $\sigma_{rp}=\prod_{k=1...N_s}\sigma_{rp}(k)$, each $\sigma_{rp}(k)$  just permuting numbers of a given site $k$, that is to say numbers $m(k-1)+1,\,\,m(k-1)+2\, ...,\,mk$ (where $1\leq k\leq {N_s}$).
They are such that:
\begin{equation}
\prod_{k=1...N_s}\sigma_{rp}(k) S_p \rightarrow S_r.
\end{equation}
To each of those $\sigma_{rp}(k)$ corresponds a linear combination of permutation  $\eta_{rp}(k)$ (involving permutation between numbers of particles of site $k$) such that:
\begin{equation}
o^{\alpha}_{r1}=\prod_{k=1...N_s}\eta_{rp}(k)o^{\alpha}_{p1}.
\end{equation}
Then, as before, since each $Proj(k)$ is equal to the orthogonal units $o^{[1 1 ...1]}_{11}$ , where $[1 1 ...1]$ is the $m$ boxes fully antisymmetric irrep, (with number $1,2,...,m$ replaced by numbers $m(k-1)+1,\,\,m(k-1)+2\, ...,\,mk$),
the properties in Eq. (\ref{relation_orthonormal}) imply that:
\begin{align}
&Proj \, o^{\alpha}_{r1}=\prod_{k=1...N_s} Proj(k) \eta_{rp}(k)o^{\alpha}_{p1} \nonumber \\
&\propto \prod_{k=1...N_s} Proj(k) o^{\alpha}_{p1}=Proj \, o^{\alpha}_{p1}.
\end{align}
This means that the number of independent linear combination of permutations generated by applying the projector on the orthogonal unit corresponding to a given class is at most 1. In addition,
it cannot be zero as a consequence of Eq. (\ref{dimension_fullHS_antisymmetric}) and (\ref{dimension_fullHS_symmetric}) (which are direct consequences of the  Itzykson-Nauenberg rules).

\subsection{Algorithm to create the subset of representative SYTs.}
\label{appendix_algo_SYTs}
For a shape $\alpha$, given as an imput, there is an algorithm  to generate all the SYTs directly in the last order sequence (iterating $f^{\alpha}$ times  the algorithm called NEXYTB in Chapter 14  of  ~Ref. \onlinecite{nexytb}),
which is very useful in the case $m=1$. 
When $m>1$ in the fully symmetric or antisymmetric irrep and with $N_s$ sites, we could proceed in the following way: for a given shape $\alpha$ (with $mN_s$ boxes), we could generate all the SYTs through the algorithm 
NEXYTB, and then select the subset which satisfies the proper internal constraints.
However, this is not efficient: for instance, for $SU(4)$, $m=2$ and $N_s=18$, in the antisymmetric case, the number of independent singlets is $\approx 61 \times 10^6$, while the total number of SYTs of shape
$[9,9,9,9]$ is $\approx 2 \times 10^{14}$.

\begin{figure}
\begin{center}
\includegraphics[width=250pt]{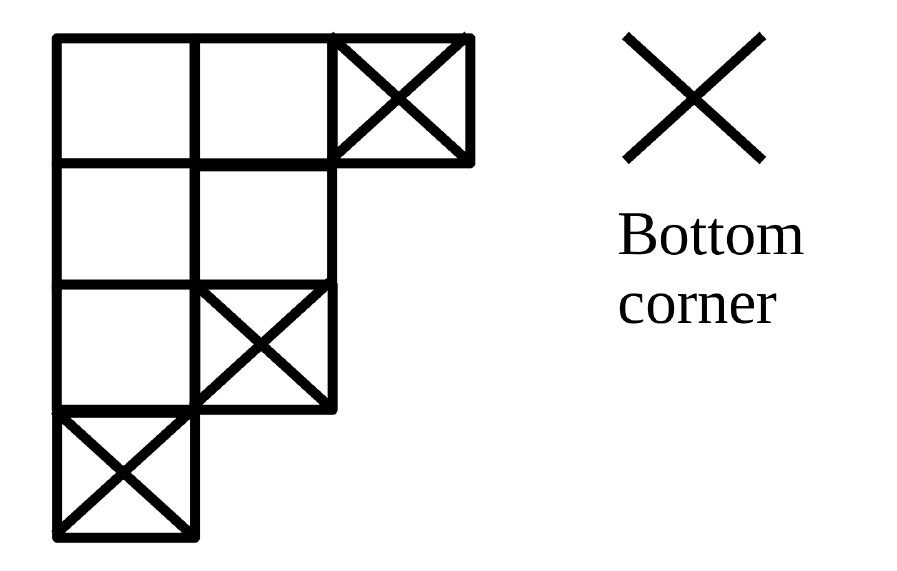}
\caption{\label{schema_bottom_corner} For a given {\it Young tableau} $\alpha=[\alpha_1,\alpha_2,...,\alpha_k]$, with $\alpha_i$ the length of the row $i$ of the shape and $k$ the number of rows of the shape, (here $\alpha=[3 2 2 1]$ and $k=4$), the bottom corners are the boxes where we could put the last number when we fill up the shape in the {\it standard way} (here, this number, which is the total number of boxes, is equal to $8$).
If we conventionnally set $\alpha_{k+1}=0$, those bottom corner correspond to all the rows $j$ such that $\alpha_j>\alpha_{j+1}$.
}
\end{center}
\end{figure}

\begin{figure}
\begin{center}
\hskip -1cm
\includegraphics[width=270pt]{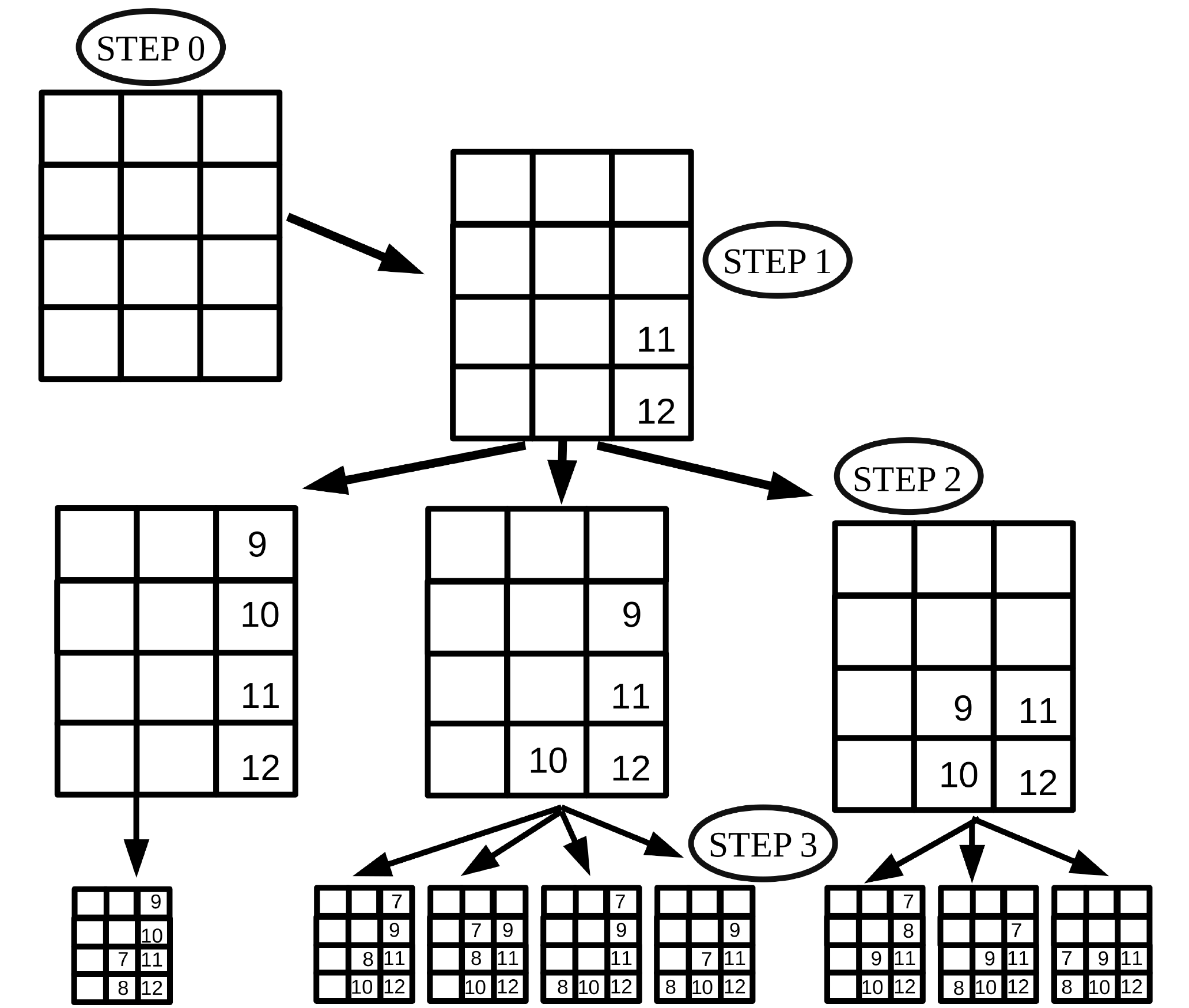}
\caption{\label{schema_generation_tbx} First three steps (out of six) to generate all the SYTs with proper internal conditions (here $y(2j-1)<y(2j)\,\,\forall j=1...6$, where $y(j)$ designates the row of the number $j$, i.e $y(11)=3$), of shape $[3333]$.
In this example, the $16$ tableaux obtained at the end of the algorithm (see text for details) represent an orthonormal basis of the $SU(4)$ singlets subspace for a $6$ sites system with 2 particles per site in the antiysmmetric representation.                                                                                                
}
\end{center}
\end{figure}

So we have devised a specific algorithm to generate directly the subset of SYTs of a given shape $\alpha$ with the proper internal constraints.
We do it for the antisymmetric case and $m=2$, the generalization to higer $m$ and the transposition to the symmetric case being straightforward.
The idea is to fill up the shapes number after number (each number labelling a particle) starting from the very last one.
In order to obtain a {\bf standard tableau}, at every stage, the number should be situated in the current {\it bottom corners} , see Fig \ref{schema_bottom_corner} for a definition.
There are as many steps to generate the standard Young tableaux (SYTs) with proper internal conditions as sites in the system, that is $N_s$ in our notation. So we perform a descending loop starting from $N_s$ and going to 1.
The purpose of each step is to fill up the current partially filled shapes with numbers of the particles of the corresponding site.
So each step is made of $m$ stages. If we start step $q>0$, the current partially filled shapes had already been filled up with numbers $mN_s, mN_s-1, ....m(N_s-q+1)+1$ from previous steps. See Fig \ref{schema_generation_tbx}, where the case $N_s=6$, $m=2$ and $\alpha=[3 3 3 3]$, (which would correspond to the creation of the $SU(4)$ singlet subspace in a $6$-sites system with 2 particles per site in the antisymmetric representation ) is treated. The purpose of step $q$ is to add to the current partially filled shapes the numbers $m(N_s-q+1),m(N_s-q+1)-1,....,m(N_s-q)+1$, in $m$ different stages.
We first locate the possible {\it bottom corners} of the current partially filled shapes for the number $m(N_s-q+1)$. One can have several possibilities. See for instance Fig \ref{schema_generation_tbx}, where for the first stage of step $2$, one must  fill the current tableau with the number $10$. The current tableau contains only $11$ and $12$, and the {\it remaining} Young tableau (the one without $11$ and $12$) has shape $[4\, 4\, 2\, 2\, 0]$ , so that there are $2$ bottom corners: one at row 2 and one at row 4.
For each possibility, one needs now to locate  $m(N_s-q+1)-1$, that is $9$ in our example. 
If our only purpose was to create all the SYTs of shape $\alpha$, it would be sufficient to put this number in the current bottom corners. But we add the {\it internal constraint}, corresponding to the local symmetry under investigation. In the antisymetric case, one also needs to have the row of $m(N_s-q+1)-1$ strictly above the one of $m(N_s-q+1)$. Denoting by $y(j)$ the row of the number $j$, one needs to have more generally $y(m(N_s-q+1))>y(m(N_s-q+1)-1)>....>y(m(N_s-q)+1)$.
That is why the step $2$ in our example shown in  Fig \ref{schema_generation_tbx} leads to the creation of $3$ (and no more) tableaux partially filled with numbers $9,10,11,$ and $12$.
We continue in this way up to $q=1$. We only need to  keep in memory the $mN_s-$dimensionnal current vectors $y$ that labels the rows of the already located numbers (we can put $0$ for numbers not located yet, that is $y(j)=0\,\, \forall\, j<m(N_s-q)+1$ if we are at step $q$.)
Interestingly, due to the additional internal constraints, the number of current partially filled shapes is not always monotically increasing with the rank of the steps: some partially filled shapes which satisfy the internal conditions for all sites between $q=j$ and $q=N_s$ might not lead to tableaux which satisfy them one step later. 
For instance, in the example shown in Fig. \ref{schema_generation_tbx}, the complete algorithm leads to $16$ differents SYTs which satisfy the internal conditions, while at the end of step 4 over 6, one already has $19$ tableaux satisfying them.

Empirically, one needs to plan a few  times more intermediate current tableaux than the number of final ones (between $1$ and $8$ for the most useful shapes).
We have used the symmetric version of this algorithm to create the $190720530$ representative SYTs of shape $[8 8 8 8]$ to create the $SU(4)$ singlets basis of a $16$ sites system with $2$ particles per site in $1$ hour and $45$ minutes with one CPU on a standard computer.

\bibliographystyle{apsrev4-1}
\bibliography{antisym_irrep_BIB}

\begin{thebibliography}{64}%
\makeatletter
\providecommand \@ifxundefined [1]{%
 \@ifx{#1\undefined}
}%
\providecommand \@ifnum [1]{%
 \ifnum #1\expandafter \@firstoftwo
 \else \expandafter \@secondoftwo
 \fi
}%
\providecommand \@ifx [1]{%
 \ifx #1\expandafter \@firstoftwo
 \else \expandafter \@secondoftwo
 \fi
}%
\providecommand \natexlab [1]{#1}%
\providecommand \enquote  [1]{``#1''}%
\providecommand \bibnamefont  [1]{#1}%
\providecommand \bibfnamefont [1]{#1}%
\providecommand \citenamefont [1]{#1}%
\providecommand \href@noop [0]{\@secondoftwo}%
\providecommand \href [0]{\begingroup \@sanitize@url \@href}%
\providecommand \@href[1]{\@@startlink{#1}\@@href}%
\providecommand \@@href[1]{\endgroup#1\@@endlink}%
\providecommand \@sanitize@url [0]{\catcode `\\12\catcode `\$12\catcode
  `\&12\catcode `\#12\catcode `\^12\catcode `\_12\catcode `\%12\relax}%
\providecommand \@@startlink[1]{}%
\providecommand \@@endlink[0]{}%
\providecommand \url  [0]{\begingroup\@sanitize@url \@url }%
\providecommand \@url [1]{\endgroup\@href {#1}{\urlprefix }}%
\providecommand \urlprefix  [0]{URL }%
\providecommand \Eprint [0]{\href }%
\providecommand \doibase [0]{http://dx.doi.org/}%
\providecommand \selectlanguage [0]{\@gobble}%
\providecommand \bibinfo  [0]{\@secondoftwo}%
\providecommand \bibfield  [0]{\@secondoftwo}%
\providecommand \translation [1]{[#1]}%
\providecommand \BibitemOpen [0]{}%
\providecommand \bibitemStop [0]{}%
\providecommand \bibitemNoStop [0]{.\EOS\space}%
\providecommand \EOS [0]{\spacefactor3000\relax}%
\providecommand \BibitemShut  [1]{\csname bibitem#1\endcsname}%
\let\auto@bib@innerbib\@empty
\bibitem [{\citenamefont {Bloch}\ \emph {et~al.}(2008)\citenamefont {Bloch},
  \citenamefont {Dalibard},\ and\ \citenamefont {Zwerger}}]{dalibard}%
  \BibitemOpen
  \bibfield  {author} {\bibinfo {author} {\bibfnamefont {I.}~\bibnamefont
  {Bloch}}, \bibinfo {author} {\bibfnamefont {J.}~\bibnamefont {Dalibard}}, \
  and\ \bibinfo {author} {\bibfnamefont {W.}~\bibnamefont {Zwerger}},\ }\href
  {\doibase 10.1103/RevModPhys.80.885} {\bibfield  {journal} {\bibinfo
  {journal} {Rev. Mod. Phys.}\ }\textbf {\bibinfo {volume} {80}},\ \bibinfo
  {pages} {885} (\bibinfo {year} {2008})}\BibitemShut {NoStop}%
\bibitem [{\citenamefont {Gorshkov}\ \emph {et~al.}(2010)\citenamefont
  {Gorshkov}, \citenamefont {Hermele}, \citenamefont {Gurarie}, \citenamefont
  {Xu}, \citenamefont {Julienne}, \citenamefont {Ye}, \citenamefont {Zoller},
  \citenamefont {Demler}, \citenamefont {Lukin},\ and\ \citenamefont
  {Rey}}]{gorshkov2010}%
  \BibitemOpen
  \bibfield  {author} {\bibinfo {author} {\bibfnamefont {A.~V.}\ \bibnamefont
  {Gorshkov}}, \bibinfo {author} {\bibfnamefont {M.}~\bibnamefont {Hermele}},
  \bibinfo {author} {\bibfnamefont {V.}~\bibnamefont {Gurarie}}, \bibinfo
  {author} {\bibfnamefont {C.}~\bibnamefont {Xu}}, \bibinfo {author}
  {\bibfnamefont {P.~S.}\ \bibnamefont {Julienne}}, \bibinfo {author}
  {\bibfnamefont {J.}~\bibnamefont {Ye}}, \bibinfo {author} {\bibfnamefont
  {P.}~\bibnamefont {Zoller}}, \bibinfo {author} {\bibfnamefont
  {E.}~\bibnamefont {Demler}}, \bibinfo {author} {\bibfnamefont {M.~D.}\
  \bibnamefont {Lukin}}, \ and\ \bibinfo {author} {\bibfnamefont {A.~M.}\
  \bibnamefont {Rey}},\ }\href {\doibase 10.1038/nphys1535} {\bibfield
  {journal} {\bibinfo  {journal} {Nat Phys}\ }\textbf {\bibinfo {volume} {6}},\
  \bibinfo {pages} {289} (\bibinfo {year} {2010})}\BibitemShut {NoStop}%
\bibitem [{\citenamefont {Taie}\ \emph {et~al.}(2012)\citenamefont {Taie},
  \citenamefont {Yamazaki}, \citenamefont {Sugawa},\ and\ \citenamefont
  {Takahashi}}]{takahashi2012}%
  \BibitemOpen
  \bibfield  {author} {\bibinfo {author} {\bibfnamefont {S.}~\bibnamefont
  {Taie}}, \bibinfo {author} {\bibfnamefont {R.}~\bibnamefont {Yamazaki}},
  \bibinfo {author} {\bibfnamefont {S.}~\bibnamefont {Sugawa}}, \ and\ \bibinfo
  {author} {\bibfnamefont {Y.}~\bibnamefont {Takahashi}},\ }\href {\doibase
  10.1038/nphys2430} {\bibfield  {journal} {\bibinfo  {journal} {Nat Phys}\
  }\textbf {\bibinfo {volume} {8}},\ \bibinfo {pages} {825} (\bibinfo {year}
  {2012})}\BibitemShut {NoStop}%
\bibitem [{\citenamefont {Pagano}\ \emph {et~al.}(2014)\citenamefont {Pagano},
  \citenamefont {Mancini}, \citenamefont {Cappellini}, \citenamefont
  {Lombardi}, \citenamefont {Sch\"{a}fer}, \citenamefont {Hu}, \citenamefont
  {Liu}, \citenamefont {Catani}, \citenamefont {Sias}, \citenamefont
  {Inguscio},\ and\ \citenamefont {Fallani}}]{pagano2014}%
  \BibitemOpen
  \bibfield  {author} {\bibinfo {author} {\bibfnamefont {G.}~\bibnamefont
  {Pagano}}, \bibinfo {author} {\bibfnamefont {M.}~\bibnamefont {Mancini}},
  \bibinfo {author} {\bibfnamefont {G.}~\bibnamefont {Cappellini}}, \bibinfo
  {author} {\bibfnamefont {P.}~\bibnamefont {Lombardi}}, \bibinfo {author}
  {\bibfnamefont {F.}~\bibnamefont {Sch\"{a}fer}}, \bibinfo {author}
  {\bibfnamefont {H.}~\bibnamefont {Hu}}, \bibinfo {author} {\bibfnamefont
  {X.-J.}\ \bibnamefont {Liu}}, \bibinfo {author} {\bibfnamefont
  {J.}~\bibnamefont {Catani}}, \bibinfo {author} {\bibfnamefont
  {C.}~\bibnamefont {Sias}}, \bibinfo {author} {\bibfnamefont {M.}~\bibnamefont
  {Inguscio}}, \ and\ \bibinfo {author} {\bibfnamefont {L.}~\bibnamefont
  {Fallani}},\ }\href {\doibase 10.1038/nphys2878} {\bibfield  {journal}
  {\bibinfo  {journal} {Nature Physics}\ }\textbf {\bibinfo {volume} {10}},\
  \bibinfo {pages} {198} (\bibinfo {year} {2014})}\BibitemShut {NoStop}%
\bibitem [{\citenamefont {{Scazza}}\ \emph {et~al.}(2014)\citenamefont
  {{Scazza}}, \citenamefont {{Hofrichter}}, \citenamefont {{H{\"o}fer}},
  \citenamefont {{De Groot}}, \citenamefont {{Bloch}},\ and\ \citenamefont
  {{F{\"o}lling}}}]{scazza}%
  \BibitemOpen
  \bibfield  {author} {\bibinfo {author} {\bibfnamefont {F.}~\bibnamefont
  {{Scazza}}}, \bibinfo {author} {\bibfnamefont {C.}~\bibnamefont
  {{Hofrichter}}}, \bibinfo {author} {\bibfnamefont {M.}~\bibnamefont
  {{H{\"o}fer}}}, \bibinfo {author} {\bibfnamefont {P.~C.}\ \bibnamefont {{De
  Groot}}}, \bibinfo {author} {\bibfnamefont {I.}~\bibnamefont {{Bloch}}}, \
  and\ \bibinfo {author} {\bibfnamefont {S.}~\bibnamefont {{F{\"o}lling}}},\
  }\href@noop {} {\bibfield  {journal} {\bibinfo  {journal} {ArXiv e-prints}\ }
  (\bibinfo {year} {2014})},\ \Eprint {http://arxiv.org/abs/1403.4761}
  {arXiv:1403.4761 [cond-mat.quant-gas]} \BibitemShut {NoStop}%
\bibitem [{\citenamefont {{Zhang}}\ \emph {et~al.}(2014)\citenamefont
  {{Zhang}}, \citenamefont {{Bishof}}, \citenamefont {{Bromley}}, \citenamefont
  {{Kraus}}, \citenamefont {{Safronova}}, \citenamefont {{Zoller}},
  \citenamefont {{Rey}},\ and\ \citenamefont {{Ye}}}]{zhang_SUN}%
  \BibitemOpen
  \bibfield  {author} {\bibinfo {author} {\bibfnamefont {X.}~\bibnamefont
  {{Zhang}}}, \bibinfo {author} {\bibfnamefont {M.}~\bibnamefont {{Bishof}}},
  \bibinfo {author} {\bibfnamefont {S.~L.}\ \bibnamefont {{Bromley}}}, \bibinfo
  {author} {\bibfnamefont {C.~V.}\ \bibnamefont {{Kraus}}}, \bibinfo {author}
  {\bibfnamefont {M.~S.}\ \bibnamefont {{Safronova}}}, \bibinfo {author}
  {\bibfnamefont {P.}~\bibnamefont {{Zoller}}}, \bibinfo {author}
  {\bibfnamefont {A.~M.}\ \bibnamefont {{Rey}}}, \ and\ \bibinfo {author}
  {\bibfnamefont {J.}~\bibnamefont {{Ye}}},\ }\href@noop {} {\bibfield
  {journal} {\bibinfo  {journal} {ArXiv e-prints}\ } (\bibinfo {year}
  {2014})},\ \Eprint {http://arxiv.org/abs/1403.2964} {arXiv:1403.2964
  [cond-mat.quant-gas]} \BibitemShut {NoStop}%
\bibitem [{\citenamefont {Sutherland}(1975)}]{sutherland}%
  \BibitemOpen
  \bibfield  {author} {\bibinfo {author} {\bibfnamefont {B.}~\bibnamefont
  {Sutherland}},\ }\href {\doibase 10.1103/PhysRevB.12.3795} {\bibfield
  {journal} {\bibinfo  {journal} {Phys. Rev. B}\ }\textbf {\bibinfo {volume}
  {12}},\ \bibinfo {pages} {3795} (\bibinfo {year} {1975})}\BibitemShut
  {NoStop}%
\bibitem [{\citenamefont {Haldane}(1983)}]{haldanegap}%
  \BibitemOpen
  \bibfield  {author} {\bibinfo {author} {\bibfnamefont {F.~D.~M.}\
  \bibnamefont {Haldane}},\ }\href {\doibase 10.1103/PhysRevLett.50.1153}
  {\bibfield  {journal} {\bibinfo  {journal} {Phys. Rev. Lett.}\ }\textbf
  {\bibinfo {volume} {50}},\ \bibinfo {pages} {1153} (\bibinfo {year}
  {1983})}\BibitemShut {NoStop}%
\bibitem [{\citenamefont {White}\ and\ \citenamefont
  {Huse}(1993)}]{whiteprb1993}%
  \BibitemOpen
  \bibfield  {author} {\bibinfo {author} {\bibfnamefont {S.~R.}\ \bibnamefont
  {White}}\ and\ \bibinfo {author} {\bibfnamefont {D.~A.}\ \bibnamefont
  {Huse}},\ }\href {\doibase 10.1103/PhysRevB.48.3844} {\bibfield  {journal}
  {\bibinfo  {journal} {Phys. Rev. B}\ }\textbf {\bibinfo {volume} {48}},\
  \bibinfo {pages} {3844} (\bibinfo {year} {1993})}\BibitemShut {NoStop}%
\bibitem [{\citenamefont {T\'oth}\ \emph {et~al.}(2010)\citenamefont {T\'oth},
  \citenamefont {L\"auchli}, \citenamefont {Mila},\ and\ \citenamefont
  {Penc}}]{toth2010}%
  \BibitemOpen
  \bibfield  {author} {\bibinfo {author} {\bibfnamefont {T.~A.}\ \bibnamefont
  {T\'oth}}, \bibinfo {author} {\bibfnamefont {A.~M.}\ \bibnamefont
  {L\"auchli}}, \bibinfo {author} {\bibfnamefont {F.}~\bibnamefont {Mila}}, \
  and\ \bibinfo {author} {\bibfnamefont {K.}~\bibnamefont {Penc}},\ }\href
  {\doibase 10.1103/PhysRevLett.105.265301} {\bibfield  {journal} {\bibinfo
  {journal} {Phys. Rev. Lett.}\ }\textbf {\bibinfo {volume} {105}},\ \bibinfo
  {pages} {265301} (\bibinfo {year} {2010})}\BibitemShut {NoStop}%
\bibitem [{\citenamefont {Bauer}\ \emph
  {et~al.}(2012{\natexlab{a}})\citenamefont {Bauer}, \citenamefont {Corboz},
  \citenamefont {Läuchli}, \citenamefont {Messio}, \citenamefont {Penc},
  \citenamefont {Troyer},\ and\ \citenamefont
  {Mila}}]{bauer_three-sublattice_2012}%
  \BibitemOpen
  \bibfield  {author} {\bibinfo {author} {\bibfnamefont {B.}~\bibnamefont
  {Bauer}}, \bibinfo {author} {\bibfnamefont {P.}~\bibnamefont {Corboz}},
  \bibinfo {author} {\bibfnamefont {A.}~\bibnamefont {Läuchli}}, \bibinfo
  {author} {\bibfnamefont {L.}~\bibnamefont {Messio}}, \bibinfo {author}
  {\bibfnamefont {K.}~\bibnamefont {Penc}}, \bibinfo {author} {\bibfnamefont
  {M.}~\bibnamefont {Troyer}}, \ and\ \bibinfo {author} {\bibfnamefont
  {F.}~\bibnamefont {Mila}},\ }\href {\doibase 10.1103/PhysRevB.85.125116}
  {\bibfield  {journal} {\bibinfo  {journal} {Physical Review B}\ }\textbf
  {\bibinfo {volume} {85}},\ \bibinfo {pages} {125116} (\bibinfo {year}
  {2012}{\natexlab{a}})}\BibitemShut {NoStop}%
\bibitem [{\citenamefont {Corboz}\ \emph {et~al.}(2011)\citenamefont {Corboz},
  \citenamefont {L\"auchli}, \citenamefont {Penc}, \citenamefont {Troyer},\
  and\ \citenamefont {Mila}}]{corbozSU42011}%
  \BibitemOpen
  \bibfield  {author} {\bibinfo {author} {\bibfnamefont {P.}~\bibnamefont
  {Corboz}}, \bibinfo {author} {\bibfnamefont {A.~M.}\ \bibnamefont
  {L\"auchli}}, \bibinfo {author} {\bibfnamefont {K.}~\bibnamefont {Penc}},
  \bibinfo {author} {\bibfnamefont {M.}~\bibnamefont {Troyer}}, \ and\ \bibinfo
  {author} {\bibfnamefont {F.}~\bibnamefont {Mila}},\ }\href {\doibase
  10.1103/PhysRevLett.107.215301} {\bibfield  {journal} {\bibinfo  {journal}
  {Phys. Rev. Lett.}\ }\textbf {\bibinfo {volume} {107}},\ \bibinfo {pages}
  {215301} (\bibinfo {year} {2011})}\BibitemShut {NoStop}%
\bibitem [{\citenamefont {Nataf}\ and\ \citenamefont {Mila}(2014)}]{nataf2014}%
  \BibitemOpen
  \bibfield  {author} {\bibinfo {author} {\bibfnamefont {P.}~\bibnamefont
  {Nataf}}\ and\ \bibinfo {author} {\bibfnamefont {F.}~\bibnamefont {Mila}},\
  }\href {\doibase 10.1103/PhysRevLett.113.127204} {\bibfield  {journal}
  {\bibinfo  {journal} {Phys. Rev. Lett.}\ }\textbf {\bibinfo {volume} {113}},\
  \bibinfo {pages} {127204} (\bibinfo {year} {2014})}\BibitemShut {NoStop}%
\bibitem [{\citenamefont {Corboz}\ \emph
  {et~al.}(2012{\natexlab{a}})\citenamefont {Corboz}, \citenamefont {Lajk\'o},
  \citenamefont {L\"auchli}, \citenamefont {Penc},\ and\ \citenamefont
  {Mila}}]{corbozPRX2012}%
  \BibitemOpen
  \bibfield  {author} {\bibinfo {author} {\bibfnamefont {P.}~\bibnamefont
  {Corboz}}, \bibinfo {author} {\bibfnamefont {M.}~\bibnamefont {Lajk\'o}},
  \bibinfo {author} {\bibfnamefont {A.~M.}\ \bibnamefont {L\"auchli}}, \bibinfo
  {author} {\bibfnamefont {K.}~\bibnamefont {Penc}}, \ and\ \bibinfo {author}
  {\bibfnamefont {F.}~\bibnamefont {Mila}},\ }\href {\doibase
  10.1103/PhysRevX.2.041013} {\bibfield  {journal} {\bibinfo  {journal} {Phys.
  Rev. X}\ }\textbf {\bibinfo {volume} {2}},\ \bibinfo {pages} {041013}
  (\bibinfo {year} {2012}{\natexlab{a}})}\BibitemShut {NoStop}%
\bibitem [{\citenamefont {Hermele}\ \emph {et~al.}(2009)\citenamefont
  {Hermele}, \citenamefont {Gurarie},\ and\ \citenamefont {Rey}}]{hermele2009}%
  \BibitemOpen
  \bibfield  {author} {\bibinfo {author} {\bibfnamefont {M.}~\bibnamefont
  {Hermele}}, \bibinfo {author} {\bibfnamefont {V.}~\bibnamefont {Gurarie}}, \
  and\ \bibinfo {author} {\bibfnamefont {A.~M.}\ \bibnamefont {Rey}},\ }\href
  {\doibase 10.1103/PhysRevLett.103.135301} {\bibfield  {journal} {\bibinfo
  {journal} {Phys. Rev. Lett.}\ }\textbf {\bibinfo {volume} {103}},\ \bibinfo
  {pages} {135301} (\bibinfo {year} {2009})}\BibitemShut {NoStop}%
\bibitem [{\citenamefont {Hermele}\ and\ \citenamefont
  {Gurarie}(2011)}]{hermele_topological_2011}%
  \BibitemOpen
  \bibfield  {author} {\bibinfo {author} {\bibfnamefont {M.}~\bibnamefont
  {Hermele}}\ and\ \bibinfo {author} {\bibfnamefont {V.}~\bibnamefont
  {Gurarie}},\ }\href {\doibase 10.1103/PhysRevB.84.174441} {\bibfield
  {journal} {\bibinfo  {journal} {Physical Review B}\ }\textbf {\bibinfo
  {volume} {84}},\ \bibinfo {pages} {1} (\bibinfo {year} {2011})}\BibitemShut
  {NoStop}%
\bibitem [{\citenamefont {Frischmuth}\ \emph {et~al.}(1999)\citenamefont
  {Frischmuth}, \citenamefont {Mila},\ and\ \citenamefont
  {Troyer}}]{Frischmuth1999}%
  \BibitemOpen
  \bibfield  {author} {\bibinfo {author} {\bibfnamefont {B.}~\bibnamefont
  {Frischmuth}}, \bibinfo {author} {\bibfnamefont {F.}~\bibnamefont {Mila}}, \
  and\ \bibinfo {author} {\bibfnamefont {M.}~\bibnamefont {Troyer}},\ }\href
  {\doibase 10.1103/PhysRevLett.82.835} {\bibfield  {journal} {\bibinfo
  {journal} {Physical Review Letters}\ }\textbf {\bibinfo {volume} {82}},\
  \bibinfo {pages} {835} (\bibinfo {year} {1999})}\BibitemShut {NoStop}%
\bibitem [{\citenamefont {Messio}\ and\ \citenamefont
  {Mila}(2012)}]{Messio2012}%
  \BibitemOpen
  \bibfield  {author} {\bibinfo {author} {\bibfnamefont {L.}~\bibnamefont
  {Messio}}\ and\ \bibinfo {author} {\bibfnamefont {F.}~\bibnamefont {Mila}},\
  }\href {\doibase 10.1103/PhysRevLett.109.205306} {\bibfield  {journal}
  {\bibinfo  {journal} {Physical Review Letters}\ }\textbf {\bibinfo {volume}
  {109}},\ \bibinfo {pages} {205306} (\bibinfo {year} {2012})}\BibitemShut
  {NoStop}%
\bibitem [{\citenamefont {Bonnes}\ \emph {et~al.}(2012)\citenamefont {Bonnes},
  \citenamefont {Hazzard}, \citenamefont {Manmana}, \citenamefont {Rey},\ and\
  \citenamefont {Wessel}}]{bonnes}%
  \BibitemOpen
  \bibfield  {author} {\bibinfo {author} {\bibfnamefont {L.}~\bibnamefont
  {Bonnes}}, \bibinfo {author} {\bibfnamefont {K.~R.~A.}\ \bibnamefont
  {Hazzard}}, \bibinfo {author} {\bibfnamefont {S.~R.}\ \bibnamefont
  {Manmana}}, \bibinfo {author} {\bibfnamefont {A.~M.}\ \bibnamefont {Rey}}, \
  and\ \bibinfo {author} {\bibfnamefont {S.}~\bibnamefont {Wessel}},\ }\href
  {\doibase 10.1103/PhysRevLett.109.205305} {\bibfield  {journal} {\bibinfo
  {journal} {Phys. Rev. Lett.}\ }\textbf {\bibinfo {volume} {109}},\ \bibinfo
  {pages} {205305} (\bibinfo {year} {2012})}\BibitemShut {NoStop}%
\bibitem [{\citenamefont {Read}\ and\ \citenamefont {Sachdev}(1989)}]{sachdev}%
  \BibitemOpen
  \bibfield  {author} {\bibinfo {author} {\bibfnamefont {N.}~\bibnamefont
  {Read}}\ and\ \bibinfo {author} {\bibfnamefont {S.}~\bibnamefont {Sachdev}},\
  }\href {\doibase http://dx.doi.org/10.1016/0550-3213(89)90061-8} {\bibfield
  {journal} {\bibinfo  {journal} {Nuclear Physics B}\ }\textbf {\bibinfo
  {volume} {316}},\ \bibinfo {pages} {609 } (\bibinfo {year}
  {1989})}\BibitemShut {NoStop}%
\bibitem [{\citenamefont {Lecheminant}\ and\ \citenamefont
  {Tsvelik}(2015)}]{lecheminant2015}%
  \BibitemOpen
  \bibfield  {author} {\bibinfo {author} {\bibfnamefont {P.}~\bibnamefont
  {Lecheminant}}\ and\ \bibinfo {author} {\bibfnamefont {A.~M.}\ \bibnamefont
  {Tsvelik}},\ }\href {\doibase 10.1103/PhysRevB.91.174407} {\bibfield
  {journal} {\bibinfo  {journal} {Phys. Rev. B}\ }\textbf {\bibinfo {volume}
  {91}},\ \bibinfo {pages} {174407} (\bibinfo {year} {2015})}\BibitemShut
  {NoStop}%
\bibitem [{\citenamefont {Nonne}\ \emph {et~al.}(2011)\citenamefont {Nonne},
  \citenamefont {Lecheminant}, \citenamefont {Capponi}, \citenamefont {Roux},\
  and\ \citenamefont {Boulat}}]{nonne2011}%
  \BibitemOpen
  \bibfield  {author} {\bibinfo {author} {\bibfnamefont {H.}~\bibnamefont
  {Nonne}}, \bibinfo {author} {\bibfnamefont {P.}~\bibnamefont {Lecheminant}},
  \bibinfo {author} {\bibfnamefont {S.}~\bibnamefont {Capponi}}, \bibinfo
  {author} {\bibfnamefont {G.}~\bibnamefont {Roux}}, \ and\ \bibinfo {author}
  {\bibfnamefont {E.}~\bibnamefont {Boulat}},\ }\href {\doibase
  10.1103/PhysRevB.84.125123} {\bibfield  {journal} {\bibinfo  {journal} {Phys.
  Rev. B}\ }\textbf {\bibinfo {volume} {84}},\ \bibinfo {pages} {125123}
  (\bibinfo {year} {2011})}\BibitemShut {NoStop}%
\bibitem [{\citenamefont {Szirmai}\ \emph {et~al.}(2011)\citenamefont
  {Szirmai}, \citenamefont {Szirmai}, \citenamefont {Zamora},\ and\
  \citenamefont {Lewenstein}}]{szirmaiSU62011}%
  \BibitemOpen
  \bibfield  {author} {\bibinfo {author} {\bibfnamefont {G.}~\bibnamefont
  {Szirmai}}, \bibinfo {author} {\bibfnamefont {E.}~\bibnamefont {Szirmai}},
  \bibinfo {author} {\bibfnamefont {A.}~\bibnamefont {Zamora}}, \ and\ \bibinfo
  {author} {\bibfnamefont {M.}~\bibnamefont {Lewenstein}},\ }\href {\doibase
  10.1103/PhysRevA.84.011611} {\bibfield  {journal} {\bibinfo  {journal} {Phys.
  Rev. A}\ }\textbf {\bibinfo {volume} {84}},\ \bibinfo {pages} {011611}
  (\bibinfo {year} {2011})}\BibitemShut {NoStop}%
\bibitem [{\citenamefont {Papanicolaou}(1988)}]{papanicolaou1988}%
  \BibitemOpen
  \bibfield  {author} {\bibinfo {author} {\bibfnamefont {N.}~\bibnamefont
  {Papanicolaou}},\ }\href {\doibase 10.1016/0550-3213(88)90073-9} {\bibfield
  {journal} {\bibinfo  {journal} {Nuclear Physics B}\ }\textbf {\bibinfo
  {volume} {305}},\ \bibinfo {pages} {367} (\bibinfo {year}
  {1988})}\BibitemShut {NoStop}%
\bibitem [{\citenamefont {Joshi}\ \emph {et~al.}(1999)\citenamefont {Joshi},
  \citenamefont {Ma}, \citenamefont {Mila}, \citenamefont {Shi},\ and\
  \citenamefont {Zhang}}]{joshi1999}%
  \BibitemOpen
  \bibfield  {author} {\bibinfo {author} {\bibfnamefont {A.}~\bibnamefont
  {Joshi}}, \bibinfo {author} {\bibfnamefont {M.}~\bibnamefont {Ma}}, \bibinfo
  {author} {\bibfnamefont {F.}~\bibnamefont {Mila}}, \bibinfo {author}
  {\bibfnamefont {D.~N.}\ \bibnamefont {Shi}}, \ and\ \bibinfo {author}
  {\bibfnamefont {F.~C.}\ \bibnamefont {Zhang}},\ }\href {\doibase
  10.1103/PhysRevB.60.6584} {\bibfield  {journal} {\bibinfo  {journal} {Phys.
  Rev. B}\ }\textbf {\bibinfo {volume} {60}},\ \bibinfo {pages} {6584}
  (\bibinfo {year} {1999})}\BibitemShut {NoStop}%
\bibitem [{\citenamefont {Beach}\ \emph {et~al.}(2009)\citenamefont {Beach},
  \citenamefont {Alet}, \citenamefont {Mambrini},\ and\ \citenamefont
  {Capponi}}]{capponiQMC}%
  \BibitemOpen
  \bibfield  {author} {\bibinfo {author} {\bibfnamefont {K.~S.~D.}\
  \bibnamefont {Beach}}, \bibinfo {author} {\bibfnamefont {F.}~\bibnamefont
  {Alet}}, \bibinfo {author} {\bibfnamefont {M.}~\bibnamefont {Mambrini}}, \
  and\ \bibinfo {author} {\bibfnamefont {S.}~\bibnamefont {Capponi}},\ }\href
  {\doibase 10.1103/PhysRevB.80.184401} {\bibfield  {journal} {\bibinfo
  {journal} {Phys. Rev. B}\ }\textbf {\bibinfo {volume} {80}},\ \bibinfo
  {pages} {184401} (\bibinfo {year} {2009})}\BibitemShut {NoStop}%
\bibitem [{\citenamefont {Assaad}(2005)}]{assaad2005}%
  \BibitemOpen
  \bibfield  {author} {\bibinfo {author} {\bibfnamefont {F.~F.}\ \bibnamefont
  {Assaad}},\ }\href {\doibase 10.1103/PhysRevB.71.075103} {\bibfield
  {journal} {\bibinfo  {journal} {Phys. Rev. B}\ }\textbf {\bibinfo {volume}
  {71}},\ \bibinfo {pages} {075103} (\bibinfo {year} {2005})}\BibitemShut
  {NoStop}%
\bibitem [{\citenamefont {Cai}\ \emph {et~al.}(2013)\citenamefont {Cai},
  \citenamefont {Hung}, \citenamefont {Wang},\ and\ \citenamefont
  {Wu}}]{cai2013}%
  \BibitemOpen
  \bibfield  {author} {\bibinfo {author} {\bibfnamefont {Z.}~\bibnamefont
  {Cai}}, \bibinfo {author} {\bibfnamefont {H.-H.}\ \bibnamefont {Hung}},
  \bibinfo {author} {\bibfnamefont {L.}~\bibnamefont {Wang}}, \ and\ \bibinfo
  {author} {\bibfnamefont {C.}~\bibnamefont {Wu}},\ }\href {\doibase
  10.1103/PhysRevB.88.125108} {\bibfield  {journal} {\bibinfo  {journal} {Phys.
  Rev. B}\ }\textbf {\bibinfo {volume} {88}},\ \bibinfo {pages} {125108}
  (\bibinfo {year} {2013})}\BibitemShut {NoStop}%
\bibitem [{\citenamefont {Lang}\ \emph {et~al.}(2013)\citenamefont {Lang},
  \citenamefont {Meng}, \citenamefont {Muramatsu}, \citenamefont {Wessel},\
  and\ \citenamefont {Assaad}}]{lang2013}%
  \BibitemOpen
  \bibfield  {author} {\bibinfo {author} {\bibfnamefont {T.~C.}\ \bibnamefont
  {Lang}}, \bibinfo {author} {\bibfnamefont {Z.~Y.}\ \bibnamefont {Meng}},
  \bibinfo {author} {\bibfnamefont {A.}~\bibnamefont {Muramatsu}}, \bibinfo
  {author} {\bibfnamefont {S.}~\bibnamefont {Wessel}}, \ and\ \bibinfo {author}
  {\bibfnamefont {F.~F.}\ \bibnamefont {Assaad}},\ }\href {\doibase
  10.1103/PhysRevLett.111.066401} {\bibfield  {journal} {\bibinfo  {journal}
  {Phys. Rev. Lett.}\ }\textbf {\bibinfo {volume} {111}},\ \bibinfo {pages}
  {066401} (\bibinfo {year} {2013})}\BibitemShut {NoStop}%
\bibitem [{\citenamefont {Zhou}\ \emph {et~al.}(2014)\citenamefont {Zhou},
  \citenamefont {Cai}, \citenamefont {Wu},\ and\ \citenamefont
  {Wang}}]{zhou2014}%
  \BibitemOpen
  \bibfield  {author} {\bibinfo {author} {\bibfnamefont {Z.}~\bibnamefont
  {Zhou}}, \bibinfo {author} {\bibfnamefont {Z.}~\bibnamefont {Cai}}, \bibinfo
  {author} {\bibfnamefont {C.}~\bibnamefont {Wu}}, \ and\ \bibinfo {author}
  {\bibfnamefont {Y.}~\bibnamefont {Wang}},\ }\href {\doibase
  10.1103/PhysRevB.90.235139} {\bibfield  {journal} {\bibinfo  {journal} {Phys.
  Rev. B}\ }\textbf {\bibinfo {volume} {90}},\ \bibinfo {pages} {235139}
  (\bibinfo {year} {2014})}\BibitemShut {NoStop}%
\bibitem [{\citenamefont {Wang}\ and\ \citenamefont
  {Vishwanath}(2009)}]{wang_z2_2009}%
  \BibitemOpen
  \bibfield  {author} {\bibinfo {author} {\bibfnamefont {F.}~\bibnamefont
  {Wang}}\ and\ \bibinfo {author} {\bibfnamefont {A.}~\bibnamefont
  {Vishwanath}},\ }\href {\doibase 10.1103/PhysRevB.80.064413} {\bibfield
  {journal} {\bibinfo  {journal} {Physical Review B}\ }\textbf {\bibinfo
  {volume} {80}},\ \bibinfo {pages} {064413} (\bibinfo {year}
  {2009})}\BibitemShut {NoStop}%
\bibitem [{\citenamefont {Paramekanti}\ and\ \citenamefont
  {Marston}(2007)}]{paramekanti_2007}%
  \BibitemOpen
  \bibfield  {author} {\bibinfo {author} {\bibfnamefont {A.}~\bibnamefont
  {Paramekanti}}\ and\ \bibinfo {author} {\bibfnamefont {J.~B.}\ \bibnamefont
  {Marston}},\ }\href {http://stacks.iop.org/0953-8984/19/i=12/a=125215}
  {\bibfield  {journal} {\bibinfo  {journal} {Journal of Physics: Condensed
  Matter}\ }\textbf {\bibinfo {volume} {19}},\ \bibinfo {pages} {125215}
  (\bibinfo {year} {2007})}\BibitemShut {NoStop}%
\bibitem [{\citenamefont {Lajko}\ and\ \citenamefont
  {Penc}(2013)}]{lajko_tetramerization_2013}%
  \BibitemOpen
  \bibfield  {author} {\bibinfo {author} {\bibfnamefont {M.}~\bibnamefont
  {Lajko}}\ and\ \bibinfo {author} {\bibfnamefont {K.}~\bibnamefont {Penc}},\
  }\href {\doibase 10.1103/PhysRevB.87.224428} {\bibfield  {journal} {\bibinfo
  {journal} {Physical Review B}\ }\textbf {\bibinfo {volume} {87}},\ \bibinfo
  {pages} {224428} (\bibinfo {year} {2013})}\BibitemShut {NoStop}%
\bibitem [{\citenamefont {Dufour}\ \emph {et~al.}(2015)\citenamefont {Dufour},
  \citenamefont {Nataf},\ and\ \citenamefont {Mila}}]{dufourPRB2015}%
  \BibitemOpen
  \bibfield  {author} {\bibinfo {author} {\bibfnamefont {J.}~\bibnamefont
  {Dufour}}, \bibinfo {author} {\bibfnamefont {P.}~\bibnamefont {Nataf}}, \
  and\ \bibinfo {author} {\bibfnamefont {F.}~\bibnamefont {Mila}},\ }\href
  {\doibase 10.1103/PhysRevB.91.174427} {\bibfield  {journal} {\bibinfo
  {journal} {Phys. Rev. B}\ }\textbf {\bibinfo {volume} {91}},\ \bibinfo
  {pages} {174427} (\bibinfo {year} {2015})}\BibitemShut {NoStop}%
\bibitem [{\citenamefont {Rachel}\ \emph {et~al.}(2009)\citenamefont {Rachel},
  \citenamefont {Thomale}, \citenamefont {Fuhringer}, \citenamefont
  {Schmitteckert},\ and\ \citenamefont {Greiter}}]{rachel2009}%
  \BibitemOpen
  \bibfield  {author} {\bibinfo {author} {\bibfnamefont {S.}~\bibnamefont
  {Rachel}}, \bibinfo {author} {\bibfnamefont {R.}~\bibnamefont {Thomale}},
  \bibinfo {author} {\bibfnamefont {M.}~\bibnamefont {Fuhringer}}, \bibinfo
  {author} {\bibfnamefont {P.}~\bibnamefont {Schmitteckert}}, \ and\ \bibinfo
  {author} {\bibfnamefont {M.}~\bibnamefont {Greiter}},\ }\href {\doibase
  10.1103/PhysRevB.80.180420} {\bibfield  {journal} {\bibinfo  {journal} {Phys.
  Rev. B}\ }\textbf {\bibinfo {volume} {80}},\ \bibinfo {pages} {180420}
  (\bibinfo {year} {2009})}\BibitemShut {NoStop}%
\bibitem [{\citenamefont {Nonne}\ \emph {et~al.}(2013)\citenamefont {Nonne},
  \citenamefont {Moliner}, \citenamefont {Capponi}, \citenamefont
  {Lecheminant},\ and\ \citenamefont {Totsuka}}]{nonne2013}%
  \BibitemOpen
  \bibfield  {author} {\bibinfo {author} {\bibfnamefont {H.}~\bibnamefont
  {Nonne}}, \bibinfo {author} {\bibfnamefont {M.}~\bibnamefont {Moliner}},
  \bibinfo {author} {\bibfnamefont {S.}~\bibnamefont {Capponi}}, \bibinfo
  {author} {\bibfnamefont {P.}~\bibnamefont {Lecheminant}}, \ and\ \bibinfo
  {author} {\bibfnamefont {K.}~\bibnamefont {Totsuka}},\ }\href
  {http://stacks.iop.org/0295-5075/102/i=3/a=37008} {\bibfield  {journal}
  {\bibinfo  {journal} {EPL (Europhysics Letters)}\ }\textbf {\bibinfo {volume}
  {102}},\ \bibinfo {pages} {37008} (\bibinfo {year} {2013})}\BibitemShut
  {NoStop}%
\bibitem [{\citenamefont {Duivenvoorden}\ and\ \citenamefont
  {Quella}(2012)}]{quella2012}%
  \BibitemOpen
  \bibfield  {author} {\bibinfo {author} {\bibfnamefont {K.}~\bibnamefont
  {Duivenvoorden}}\ and\ \bibinfo {author} {\bibfnamefont {T.}~\bibnamefont
  {Quella}},\ }\href {\doibase 10.1103/PhysRevB.86.235142} {\bibfield
  {journal} {\bibinfo  {journal} {Phys. Rev. B}\ }\textbf {\bibinfo {volume}
  {86}},\ \bibinfo {pages} {235142} (\bibinfo {year} {2012})}\BibitemShut
  {NoStop}%
\bibitem [{\citenamefont {Fuhringer}\ \emph {et~al.}(2008)\citenamefont
  {Fuhringer}, \citenamefont {Rachel}, \citenamefont {Thomale}, \citenamefont
  {Greiter},\ and\ \citenamefont {Schmitteckert}}]{fuhringer2008}%
  \BibitemOpen
  \bibfield  {author} {\bibinfo {author} {\bibfnamefont {M.}~\bibnamefont
  {Fuhringer}}, \bibinfo {author} {\bibfnamefont {S.}~\bibnamefont {Rachel}},
  \bibinfo {author} {\bibfnamefont {R.}~\bibnamefont {Thomale}}, \bibinfo
  {author} {\bibfnamefont {M.}~\bibnamefont {Greiter}}, \ and\ \bibinfo
  {author} {\bibfnamefont {P.}~\bibnamefont {Schmitteckert}},\ }\href {\doibase
  10.1002/andp.200810326} {\bibfield  {journal} {\bibinfo  {journal} {Annalen
  der Physik}\ }\textbf {\bibinfo {volume} {17}},\ \bibinfo {pages} {922}
  (\bibinfo {year} {2008})}\BibitemShut {NoStop}%
\bibitem [{\citenamefont {Manmana}\ \emph {et~al.}(2011)\citenamefont
  {Manmana}, \citenamefont {Hazzard}, \citenamefont {Chen}, \citenamefont
  {Feiguin},\ and\ \citenamefont {Rey}}]{manmana2011}%
  \BibitemOpen
  \bibfield  {author} {\bibinfo {author} {\bibfnamefont {S.~R.}\ \bibnamefont
  {Manmana}}, \bibinfo {author} {\bibfnamefont {K.~R.~A.}\ \bibnamefont
  {Hazzard}}, \bibinfo {author} {\bibfnamefont {G.}~\bibnamefont {Chen}},
  \bibinfo {author} {\bibfnamefont {A.~E.}\ \bibnamefont {Feiguin}}, \ and\
  \bibinfo {author} {\bibfnamefont {A.~M.}\ \bibnamefont {Rey}},\ }\href
  {\doibase 10.1103/PhysRevA.84.043601} {\bibfield  {journal} {\bibinfo
  {journal} {Phys. Rev. A}\ }\textbf {\bibinfo {volume} {84}},\ \bibinfo
  {pages} {043601} (\bibinfo {year} {2011})}\BibitemShut {NoStop}%
\bibitem [{\citenamefont {Bauer}\ \emph
  {et~al.}(2012{\natexlab{b}})\citenamefont {Bauer}, \citenamefont {Corboz},
  \citenamefont {L\"auchli}, \citenamefont {Messio}, \citenamefont {Penc},
  \citenamefont {Troyer},\ and\ \citenamefont {Mila}}]{bauer2012}%
  \BibitemOpen
  \bibfield  {author} {\bibinfo {author} {\bibfnamefont {B.}~\bibnamefont
  {Bauer}}, \bibinfo {author} {\bibfnamefont {P.}~\bibnamefont {Corboz}},
  \bibinfo {author} {\bibfnamefont {A.~M.}\ \bibnamefont {L\"auchli}}, \bibinfo
  {author} {\bibfnamefont {L.}~\bibnamefont {Messio}}, \bibinfo {author}
  {\bibfnamefont {K.}~\bibnamefont {Penc}}, \bibinfo {author} {\bibfnamefont
  {M.}~\bibnamefont {Troyer}}, \ and\ \bibinfo {author} {\bibfnamefont
  {F.}~\bibnamefont {Mila}},\ }\href {\doibase 10.1103/PhysRevB.85.125116}
  {\bibfield  {journal} {\bibinfo  {journal} {Phys. Rev. B}\ }\textbf {\bibinfo
  {volume} {85}},\ \bibinfo {pages} {125116} (\bibinfo {year}
  {2012}{\natexlab{b}})}\BibitemShut {NoStop}%
\bibitem [{\citenamefont {Corboz}\ \emph
  {et~al.}(2012{\natexlab{b}})\citenamefont {Corboz}, \citenamefont {Penc},
  \citenamefont {Mila},\ and\ \citenamefont {L\"auchli}}]{corbozsimplex2012}%
  \BibitemOpen
  \bibfield  {author} {\bibinfo {author} {\bibfnamefont {P.}~\bibnamefont
  {Corboz}}, \bibinfo {author} {\bibfnamefont {K.}~\bibnamefont {Penc}},
  \bibinfo {author} {\bibfnamefont {F.}~\bibnamefont {Mila}}, \ and\ \bibinfo
  {author} {\bibfnamefont {A.~M.}\ \bibnamefont {L\"auchli}},\ }\href {\doibase
  10.1103/PhysRevB.86.041106} {\bibfield  {journal} {\bibinfo  {journal} {Phys.
  Rev. B}\ }\textbf {\bibinfo {volume} {86}},\ \bibinfo {pages} {041106}
  (\bibinfo {year} {2012}{\natexlab{b}})}\BibitemShut {NoStop}%
\bibitem [{\citenamefont {Itzykson}\ and\ \citenamefont
  {Nauenberg}(1966)}]{itzykson}%
  \BibitemOpen
  \bibfield  {author} {\bibinfo {author} {\bibfnamefont {C.}~\bibnamefont
  {Itzykson}}\ and\ \bibinfo {author} {\bibfnamefont {M.}~\bibnamefont
  {Nauenberg}},\ }\href {\doibase 10.1103/RevModPhys.38.95} {\bibfield
  {journal} {\bibinfo  {journal} {Rev. Mod. Phys.}\ }\textbf {\bibinfo {volume}
  {38}},\ \bibinfo {pages} {95} (\bibinfo {year} {1966})}\BibitemShut {NoStop}%
\bibitem [{Note1()}]{Note1}%
  \BibitemOpen
  \bibinfo {note} {In paragraph \ref {symmetric_classes}, we give a
  mathematical definition of those numbers in terms of {\protect \it Kostka}
  numbers}\BibitemShut {NoStop}%
\bibitem [{\citenamefont {Stanley}(1999)}]{stanley}%
  \BibitemOpen
  \bibfield  {author} {\bibinfo {author} {\bibfnamefont {R.~P.}\ \bibnamefont
  {Stanley}},\ }\href@noop {} {\emph {\bibinfo {title} {Enumerative
  Combinatorics}}}\ (\bibinfo  {publisher} {Cambridge University Press},\
  \bibinfo {year} {1999})\BibitemShut {NoStop}%
\bibitem [{\citenamefont {Affleck}(1986{\natexlab{a}})}]{Affleck1986a}%
  \BibitemOpen
  \bibfield  {author} {\bibinfo {author} {\bibfnamefont {I.}~\bibnamefont
  {Affleck}},\ }\href
  {http://www.sciencedirect.com/science/article/pii/0550321386901677}
  {\bibfield  {journal} {\bibinfo  {journal} {Nuclear Physics B}\ }\textbf
  {\bibinfo {volume} {265}},\ \bibinfo {pages} {409} (\bibinfo {year}
  {1986}{\natexlab{a}})}\BibitemShut {NoStop}%
\bibitem [{\citenamefont {Affleck}(1988)}]{Affleck1988}%
  \BibitemOpen
  \bibfield  {author} {\bibinfo {author} {\bibfnamefont {I.}~\bibnamefont
  {Affleck}},\ }\href
  {http://www.sciencedirect.com/science/article/pii/0550321388901174}
  {\bibfield  {journal} {\bibinfo  {journal} {Nuclear Physics B}\ }\textbf
  {\bibinfo {volume} {305}},\ \bibinfo {pages} {582} (\bibinfo {year}
  {1988})}\BibitemShut {NoStop}%
\bibitem [{Note2()}]{Note2}%
  \BibitemOpen
  \bibinfo {note} {For periodic boundary conditions, we have indexed the sites
  in such a way that the difference between two connected sites is at most 2 by
  starting from 1 at some site ({\protect \it the center}) and locating
  consecutive numbers alternatively to the left and to the right of this
  {\protect \it center}.}\BibitemShut {Stop}%
\bibitem [{\citenamefont {Affleck}(1986{\natexlab{b}})}]{affleck1986}%
  \BibitemOpen
  \bibfield  {author} {\bibinfo {author} {\bibfnamefont {I.}~\bibnamefont
  {Affleck}},\ }\href {\doibase 10.1103/PhysRevLett.56.746} {\bibfield
  {journal} {\bibinfo  {journal} {Phys. Rev. Lett.}\ }\textbf {\bibinfo
  {volume} {56}},\ \bibinfo {pages} {746} (\bibinfo {year}
  {1986}{\natexlab{b}})}\BibitemShut {NoStop}%
\bibitem [{\citenamefont {Bl\"ote}\ \emph {et~al.}(1986)\citenamefont
  {Bl\"ote}, \citenamefont {Cardy},\ and\ \citenamefont
  {Nightingale}}]{bloteprl56}%
  \BibitemOpen
  \bibfield  {author} {\bibinfo {author} {\bibfnamefont {H.~W.~J.}\
  \bibnamefont {Bl\"ote}}, \bibinfo {author} {\bibfnamefont {J.~L.}\
  \bibnamefont {Cardy}}, \ and\ \bibinfo {author} {\bibfnamefont {M.~P.}\
  \bibnamefont {Nightingale}},\ }\href {\doibase 10.1103/PhysRevLett.56.742}
  {\bibfield  {journal} {\bibinfo  {journal} {Phys. Rev. Lett.}\ }\textbf
  {\bibinfo {volume} {56}},\ \bibinfo {pages} {742} (\bibinfo {year}
  {1986})}\BibitemShut {NoStop}%
\bibitem [{\citenamefont {Moreo}(1987)}]{moreo}%
  \BibitemOpen
  \bibfield  {author} {\bibinfo {author} {\bibfnamefont {A.}~\bibnamefont
  {Moreo}},\ }\href {\doibase 10.1103/PhysRevB.36.8582} {\bibfield  {journal}
  {\bibinfo  {journal} {Phys. Rev. B}\ }\textbf {\bibinfo {volume} {36}},\
  \bibinfo {pages} {8582} (\bibinfo {year} {1987})}\BibitemShut {NoStop}%
\bibitem [{\citenamefont {Cardy}(1984)}]{cardy}%
  \BibitemOpen
  \bibfield  {author} {\bibinfo {author} {\bibfnamefont {J.~L.}\ \bibnamefont
  {Cardy}},\ }\href {http://stacks.iop.org/0305-4470/17/i=7/a=003} {\bibfield
  {journal} {\bibinfo  {journal} {Journal of Physics A: Mathematical and
  General}\ }\textbf {\bibinfo {volume} {17}},\ \bibinfo {pages} {L385}
  (\bibinfo {year} {1984})}\BibitemShut {NoStop}%
\bibitem [{\citenamefont {Lecheminant}(2015)}]{Lecheminant2015b}%
  \BibitemOpen
  \bibfield  {author} {\bibinfo {author} {\bibfnamefont {P.}~\bibnamefont
  {Lecheminant}},\ }\href {\doibase
  http://dx.doi.org/10.1016/j.nuclphysb.2015.11.004} {\bibfield  {journal}
  {\bibinfo  {journal} {Nuclear Physics B}\ }\textbf {\bibinfo {volume}
  {901}},\ \bibinfo {pages} {510 } (\bibinfo {year} {2015})}\BibitemShut
  {NoStop}%
\bibitem [{\citenamefont {Andrei}\ and\ \citenamefont
  {Johannesson}(1984)}]{andrei1984}%
  \BibitemOpen
  \bibfield  {author} {\bibinfo {author} {\bibfnamefont {N.}~\bibnamefont
  {Andrei}}\ and\ \bibinfo {author} {\bibfnamefont {H.}~\bibnamefont
  {Johannesson}},\ }\href {\doibase
  http://dx.doi.org/10.1016/0375-9601(84)90819-3} {\bibfield  {journal}
  {\bibinfo  {journal} {Physics Letters A}\ }\textbf {\bibinfo {volume}
  {104}},\ \bibinfo {pages} {370 } (\bibinfo {year} {1984})}\BibitemShut
  {NoStop}%
\bibitem [{\citenamefont {Ziman}\ and\ \citenamefont
  {Schulz}(1987)}]{ziman1987}%
  \BibitemOpen
  \bibfield  {author} {\bibinfo {author} {\bibfnamefont {T.}~\bibnamefont
  {Ziman}}\ and\ \bibinfo {author} {\bibfnamefont {H.~J.}\ \bibnamefont
  {Schulz}},\ }\href {\doibase 10.1103/PhysRevLett.59.140} {\bibfield
  {journal} {\bibinfo  {journal} {Phys. Rev. Lett.}\ }\textbf {\bibinfo
  {volume} {59}},\ \bibinfo {pages} {140} (\bibinfo {year} {1987})}\BibitemShut
  {NoStop}%
\bibitem [{\citenamefont {Weichselbaum}(2012)}]{weichselbaum2012}%
  \BibitemOpen
  \bibfield  {author} {\bibinfo {author} {\bibfnamefont {A.}~\bibnamefont
  {Weichselbaum}},\ }\href {\doibase
  http://dx.doi.org/10.1016/j.aop.2012.07.009} {\bibfield  {journal} {\bibinfo
  {journal} {Annals of Physics}\ }\textbf {\bibinfo {volume} {327}},\ \bibinfo
  {pages} {2972 } (\bibinfo {year} {2012})}\BibitemShut {NoStop}%
\bibitem [{\citenamefont {McCulloch}(2007)}]{mcculloch2007}%
  \BibitemOpen
  \bibfield  {author} {\bibinfo {author} {\bibfnamefont {I.~P.}\ \bibnamefont
  {McCulloch}},\ }\href {http://stacks.iop.org/1742-5468/2007/i=10/a=P10014}
  {\bibfield  {journal} {\bibinfo  {journal} {Journal of Statistical Mechanics:
  Theory and Experiment}\ }\textbf {\bibinfo {volume} {2007}},\ \bibinfo
  {pages} {P10014} (\bibinfo {year} {2007})}\BibitemShut {NoStop}%
\bibitem [{\citenamefont {McCulloch}\ and\ \citenamefont
  {Gulácsi}(2002)}]{mcculloch2002}%
  \BibitemOpen
  \bibfield  {author} {\bibinfo {author} {\bibfnamefont {I.~P.}\ \bibnamefont
  {McCulloch}}\ and\ \bibinfo {author} {\bibfnamefont {M.}~\bibnamefont
  {Gulácsi}},\ }\href {http://stacks.iop.org/0295-5075/57/i=6/a=852}
  {\bibfield  {journal} {\bibinfo  {journal} {EPL (Europhysics Letters)}\
  }\textbf {\bibinfo {volume} {57}},\ \bibinfo {pages} {852} (\bibinfo {year}
  {2002})}\BibitemShut {NoStop}%
\bibitem [{\citenamefont {Singh}\ \emph {et~al.}(2010)\citenamefont {Singh},
  \citenamefont {Pfeifer},\ and\ \citenamefont {Vidal}}]{gvidal2010}%
  \BibitemOpen
  \bibfield  {author} {\bibinfo {author} {\bibfnamefont {S.}~\bibnamefont
  {Singh}}, \bibinfo {author} {\bibfnamefont {R.~N.~C.}\ \bibnamefont
  {Pfeifer}}, \ and\ \bibinfo {author} {\bibfnamefont {G.}~\bibnamefont
  {Vidal}},\ }\href {\doibase 10.1103/PhysRevA.82.050301} {\bibfield  {journal}
  {\bibinfo  {journal} {Phys. Rev. A}\ }\textbf {\bibinfo {volume} {82}},\
  \bibinfo {pages} {050301} (\bibinfo {year} {2010})}\BibitemShut {NoStop}%
\bibitem [{\citenamefont {Greiter}\ and\ \citenamefont
  {Rachel}(2007)}]{rachel2007}%
  \BibitemOpen
  \bibfield  {author} {\bibinfo {author} {\bibfnamefont {M.}~\bibnamefont
  {Greiter}}\ and\ \bibinfo {author} {\bibfnamefont {S.}~\bibnamefont
  {Rachel}},\ }\href {\doibase 10.1103/PhysRevB.75.184441} {\bibfield
  {journal} {\bibinfo  {journal} {Phys. Rev. B}\ }\textbf {\bibinfo {volume}
  {75}},\ \bibinfo {pages} {184441} (\bibinfo {year} {2007})}\BibitemShut
  {NoStop}%
\bibitem [{\citenamefont {Morimoto}\ \emph {et~al.}(2014)\citenamefont
  {Morimoto}, \citenamefont {Ueda}, \citenamefont {Momoi},\ and\ \citenamefont
  {Furusaki}}]{morimoto2014}%
  \BibitemOpen
  \bibfield  {author} {\bibinfo {author} {\bibfnamefont {T.}~\bibnamefont
  {Morimoto}}, \bibinfo {author} {\bibfnamefont {H.}~\bibnamefont {Ueda}},
  \bibinfo {author} {\bibfnamefont {T.}~\bibnamefont {Momoi}}, \ and\ \bibinfo
  {author} {\bibfnamefont {A.}~\bibnamefont {Furusaki}},\ }\href {\doibase
  10.1103/PhysRevB.90.235111} {\bibfield  {journal} {\bibinfo  {journal} {Phys.
  Rev. B}\ }\textbf {\bibinfo {volume} {90}},\ \bibinfo {pages} {235111}
  (\bibinfo {year} {2014})}\BibitemShut {NoStop}%
\bibitem [{\citenamefont {Duivenvoorden}\ and\ \citenamefont
  {Quella}(2013)}]{quella2013}%
  \BibitemOpen
  \bibfield  {author} {\bibinfo {author} {\bibfnamefont {K.}~\bibnamefont
  {Duivenvoorden}}\ and\ \bibinfo {author} {\bibfnamefont {T.}~\bibnamefont
  {Quella}},\ }\href {\doibase 10.1103/PhysRevB.87.125145} {\bibfield
  {journal} {\bibinfo  {journal} {Phys. Rev. B}\ }\textbf {\bibinfo {volume}
  {87}},\ \bibinfo {pages} {125145} (\bibinfo {year} {2013})}\BibitemShut
  {NoStop}%
\bibitem [{\citenamefont {Arovas}(2008)}]{arovas2008}%
  \BibitemOpen
  \bibfield  {author} {\bibinfo {author} {\bibfnamefont {D.~P.}\ \bibnamefont
  {Arovas}},\ }\href {\doibase 10.1103/PhysRevB.77.104404} {\bibfield
  {journal} {\bibinfo  {journal} {Phys. Rev. B}\ }\textbf {\bibinfo {volume}
  {77}},\ \bibinfo {pages} {104404} (\bibinfo {year} {2008})}\BibitemShut
  {NoStop}%
\bibitem [{\citenamefont {Rutherford}(1948)}]{rutherford}%
  \BibitemOpen
  \bibfield  {author} {\bibinfo {author} {\bibfnamefont {D.~E.}\ \bibnamefont
  {Rutherford}},\ }\href@noop {} {\emph {\bibinfo {title} {Substitutional
  Analysis}}}\ (\bibinfo  {publisher} {Edinburgh University Press},\ \bibinfo
  {year} {1948})\BibitemShut {NoStop}%
\bibitem [{\citenamefont {Nijenhuis}\ and\ \citenamefont
  {Wilf}(1978)}]{nexytb}%
  \BibitemOpen
  \bibfield  {author} {\bibinfo {author} {\bibfnamefont {A.}~\bibnamefont
  {Nijenhuis}}\ and\ \bibinfo {author} {\bibfnamefont {H.~S.}\ \bibnamefont
  {Wilf}},\ }\href@noop {} {\emph {\bibinfo {title} {Combinatorial
  Algorithms}}}\ (\bibinfo  {publisher} {Academic Press New York},\ \bibinfo
  {year} {1978})\BibitemShut {NoStop}%
\end{thebibliography}%

  \end{document}